\documentclass[%
 reprint,
 amsmath,amssymb,
 aps,
prb,
]{revtex4-2}

\usepackage{graphicx}
\usepackage{dcolumn}
\usepackage{bm}
\usepackage{braket}
\usepackage{changes}

\usepackage{comment}
\usepackage{hyperref}       
\hypersetup{
    colorlinks=true,
    linkcolor=blue,
    filecolor=magenta,      
    urlcolor=blue,
}
\usepackage{color,soul}

\begin{document}

\preprint{APS/1}

\title{Topological dissipative Kerr soliton combs in a valley photonic crystal resonator}

\author{Zhen Jiang$^{1,2}$}%
\author{Lefeng Zhou$^{1}$}%
\author{Wei Li$^{3}$}%
\author{Yudong Li$^{3}$}%
\author{Liangsen Feng$^{3}$}%
\author{Tengfei Wu$^{3}$}%
\author{Chun Jiang$^{1}$}%
\email{cjiang@sjtu.edu.cn}	
\author{Guangqiang He$^{1,2}$}%
\email{gqhe@sjtu.edu.cn}	
\affiliation{%
 $^1$State Key Laboratory of Advanced Optical Communication Systems and Networks, Department of Electronic Engineering, Shanghai Jiao Tong University, Shanghai 200240, China\\
 $^2$SJTU Pinghu Institute of Intelligent Optoelectronics, Department of Electronic Engineering, Shanghai Jiao Tong University, Shanghai 200240, China \\
 $^3$Science and Technology on Metrology and Calibration Laboratory, Changcheng Institute of Metrology $\rm \&$ Measurement, Aviation Industry Corporation of China, Beijing 100095, China
}%
	
\date{\today}


\maketitle

\textbf{Topological phases have become an enabling role in exploiting new applications of nonlinear optics in recent years. Here we theoretically propose a valley photonic crystal resonator emulating topologically protected dissipative Kerr soliton combs. It is shown that topological resonator modes can be observed in the resonator. Moreover, we also simulate the dynamic evolution of the topological resonator with the injection of a continuous-wave pump laser. We find that the topological optical frequency combs evolve from Turing rolls to chaotic states, and eventually into single soliton states. More importantly, such dissipative Kerr soliton combs generated in the resonator are inborn topologically protected, showing robustness against sharp bends and structural disorders. Our design supporting topologically protected dissipative Kerr soliton combs could be implemented experimentally in on-chip nanofabricated photonic devices.}

\section{Introduction}
Optical frequency combs have been undergoing revolutionized development in integrated photonic resonators. For coherent optical frequency combs in nonlinear micro-resonators, there exist stationary temporal solutions so-called dissipative Kerr solitons (DKSs). Such DKSs are attributed to the double balance between the micro-resonators Kerr effect and dispersion management, as well as losses and parametric gain~\cite{1}. This specific phenomenon of optical frequency combs has exploited numerous concepts, such as the Stokes soliton~\cite{2}, breathing solitons~\cite{3}, and soliton crystals~\cite{4}. Primitively, bulk crystalline and microdisks~\cite{5,6,7} are proved to be an appropriate breeding ground for DKS combs. Furthermore, the soliton combs are expanded to on-chip photonic devices including $\rm Si_{3}N_{4}$~\cite{8,9,10}, $\rm LiNbO_3$~\cite{11,12}, and AlGaAs~\cite{13}.

\begin{figure*}
\centering
\includegraphics[width=0.7\textwidth]{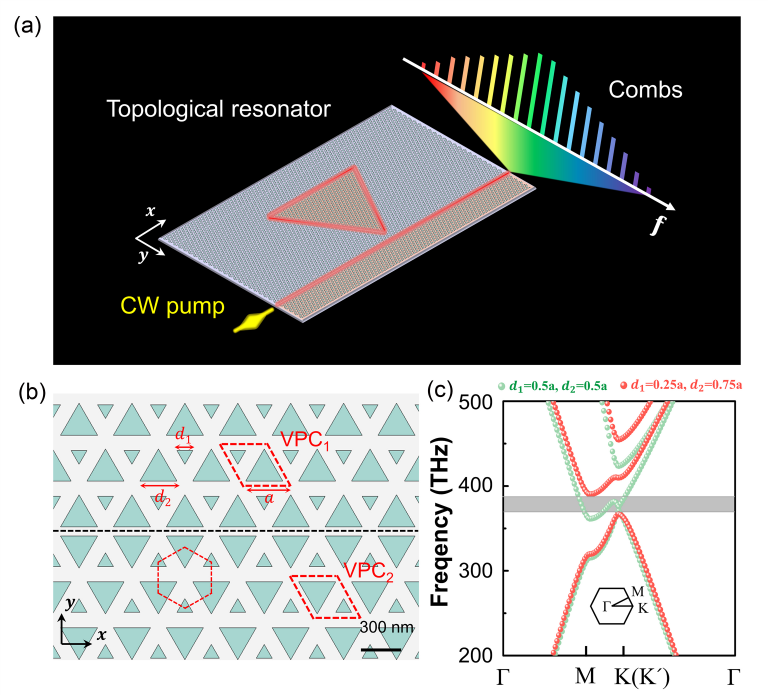}
\caption{(a) A scheme of VPC topological resonators enforcing the DKS combs. (b) 2D close-up image of VPCs with lattice constant $a=300$ nm. (c) Calculated band structures of different VPCs, where the undeformed unit cells ($d_1=d_2=0.5a$) and deformed unit cells ($d_1=0.25a$, $d_2=0.75a$) are denoted by green and red dots respectively. }

    \label{fig:1}
\end{figure*}

At the same time, topological phases of matter bring robustness to photonic devices. In parallel, advances in topological phases also excite sparks in nonlinear photonic systems. It has been experimentally proposed that the topological transport of second or third-harmonic waves can be conducted in chip-scale devices~\cite{14,15}. Furthermore, such systems exhibit the topological protection of four-wave mixing processes~\cite{16}. Lately, theoretical research demonstrates topological optical frequency combs and temporal DKSs can be manipulated in coupled micro-resonators~\cite{17}. Topological nonlinear optics also gives inspiration for implementing topological protection of complex nonlinear processes, including topological exciton-polaritons~\cite{18,19} and gap solitons~\cite{20,21}.
The topological platforms are also applied to build topological protection of quantum states, such as topological quantum sources~\cite{22}, entangled states~\cite{23,24,25}, and quantum interference~\cite{26}. Recent research has proposed a topological phase, namely the quantum valley Hall (QVH) effect~\cite{27,28,29,30,31}, to implement valley kink states along the topological interface.

Here we theoretically investigate the generation and topological transport of DKS combs in a valley photonic crystal (VPC) slab. We confirm the existence of topologically protected kink states at the resonator modes in a whisper-gallery resonator. The nonlinear dynamic evolution of the topological resonator with an excited continuous-wave (CW) pump laser is numerically simulated. The result reveals that with the detuning of the pump laser, the optical frequency combs generated in the resonator can evolve into single soliton combs eventually. More importantly, these DKSs generated in the VPC resonator are topologically protected and show robustness against sharp bends and structural imperfections. Our design may give a new spark to the chip-scale operation of topological optical frequency combs.

\section{Topological resonators}

Topological resonator is a burgeoning platform integrating topological phases and cavity dynamics. Here we demonstrate a triangle topological resonator enforcing the DKS combs in a silicon nitride ($\rm Si_{3}N_{4}$) VPC slab. As depicted in Fig.~\ref{fig:1}(a), a scheme of the VPC topological resonator is designed to produce topological frequency combs, where a bus waveguide is used to couple the pump into the resonator. When a CW laser is pumped at a resonator frequency, the third-order nonlinearity of $\rm Si_{3}N_{4}$ material with appropriate dispersion leads to a frequency comb with the spacing of free spectral range (FSR). When it meets a balance between intrinsic dispersion and nonlinearity-induced parametric gain, the comb shows a fully coherent formation, so-called DKSs. Since the DKS comb is produced in the triangle topological resonator, it is inborn topological. Such topological protection brings the DKS combs robustness against sharp corners.

The two-dimensional (2D) close-up image of VPCs is shown in Fig.~\ref{fig:1}(b), which is composed of $\rm VPC_1$ and $\rm VPC_2$~\cite{28,29}. The nanohole sizes are determined by $d_1$ and $d_2$, where the lattice constant is $a=300$ nm. As shown in Fig.~\ref{fig:1}(c), the calculated band structures of different VPCs reveal that there is a Dirac cone at the $\rm K$ and $\rm K'$ valleys as illustrated by green dots. The equilateral triangular nanoholes with $C_6$ lattice symmetry preserve a graphene-like lattice model~\cite{29}. This lattice symmetry leads to degenerate Dirac points at the $\rm K$ and $\rm K'$ valleys in the Brillouin zone. With the distortion of the unit cell ($d_1=0.25a$, and $d_2=0.75a$), it opens a photonic bandgap at the Dirac cone due to the breaking of lattice symmetry, as displayed by red dots in Fig.~\ref{fig:1}(c). Theoretically, the valley Chern numbers of  $\rm VPC_1$ and $\rm VPC_2$ are described by  $\rm C_{K⁄K'}=\pm1⁄2$, respectively (Supplemental Material). This model exhibits valley kink states propagating along the interface between $\rm VPC_1$ and $\rm VPC_2$, which shows robustness against certain disorders.

\begin{figure*}
\centering
\includegraphics[width=0.7\textwidth]{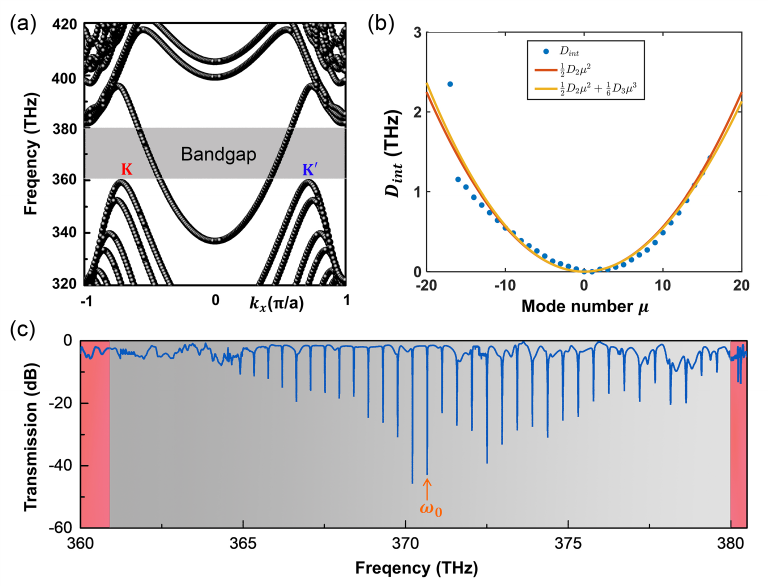}
\caption{(a) Calculated dispersion relation of valley kink states composed of $\rm VPC_1$ and $\rm VPC_2$. (b) Dispersions of the topological resonator. (c) Transmission spectrum of kink states for the topological resonator. }
    \label{fig:2}
\end{figure*}

 To study the underlying properties of topological VPCs, we perform the band calculation of VPCs. As shown in Fig.~\ref{fig:2}(a), the band opens a large gap from 361 THz to 380 THz. There exist a pair of valley-polarized topological kink states (corresponding to $k_x>0$ and $k_x<0$ respectively) in the complete bandgap. Due to the QVH effect, these kink states are locked to different valleys respectively, which refers to “valley-locked” chirality. Accordingly, this allows opposite propagating modes for kink states. These valley kink states are proven to be robust against defects and sharp corners (Supplemental Material). In addition, the light confinement of propagating modes along the interface results in the high-efficiency generation of frequency combs.
For the resonators producing the optical frequency combs, the developed dispersion engineering, so-called integrated dispersion, plays an important role in comb formation~\cite{1}. A mode number $\mu$ is used to index the relative resonator mode counted concerning the pump mode $\omega_0$. The Taylor expansion for indexed resonance modes $\omega_\mu$ around $\omega_0$ is given by 

\begin{equation}
\omega_{\mu}=\omega_{0}+\mu D_{1}+\frac{D_{2}}{2}\mu^{2}+\frac{D_{3}}{6}\mu^{3}=\omega_{0}+\sum_{i}^{\infty}D_{i}\frac{\mu^{i}}{i!},
\label{eq:1}
\end{equation}
where the expansion term is described as $D_{i}=d^{i}\omega_{\mu}/d\mu^{i}$ at $\omega=\omega_{0}$. The first-order term $D_1$ is related to the FSR of the resonator, which can be calculated as $D_{1}=\Delta\omega_{FSR}=(\omega_{1}-\omega_{-1})/2$. And the second-order dispersion term $D_2$ corresponds to the group velocity dispersion (GVD): $D_2=\omega_1+\omega_{-1}-2\omega_0$. Furthermore, it is convenient to define the integrated dispersion $D_{int}$ as

\begin{equation}
D_{int}(\mu)=\omega_{\mu}-(\omega_{0}+\mu D_{1})=\frac{D_{2}}{2}\mu^{2}+\frac{D_{3}}{6}\mu^{3}+\cdots.
\label{eq:2}
\end{equation}

Integrated dispersion $D_{int}$ contains high-order dispersion terms of the resonator and can be calculated from the transmission spectrum. Fig.~\ref{fig:2}(b) shows dispersions of the topological resonator extracted from the simulated transmission (Fig.~\ref{fig:2}(c)), which indicates an anomalous GVD case with $D_2>0$. Anomalous GVD is crucial for the generation of parametric oscillations and Kerr solitons.

\begin{figure*}
\centering
\includegraphics[width=1\textwidth]{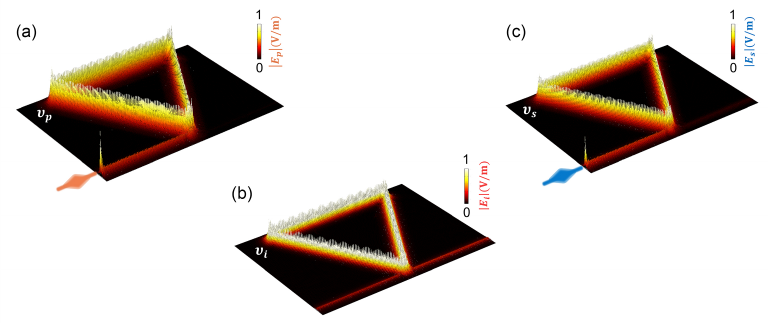}
\caption{Field profiles of the stimulated FWM process in the topological resonator at  frequencies of the (a) pump ($v_{p}=370.67$ THz), (b) idler ($v_i=370.21$ THz), and (c) signal ($v_s=371.13$ THz) respectively.}
    \label{fig:3}
\end{figure*}

Topological resonators based on VPCs have been proposed to conduct the lasing~\cite{32,33}, optical routing~\cite{34}, filters~\cite{35}, and quantum emitter~\cite{36}. Here we proposed a triangle topological resonator to produce DKS combs. A straight topological bus waveguide is used to couple the pump mode into the topological resonator, and then, guide the generated combs from the resonator to the bus waveguide. The gap width between the bus waveguide and the topological resonator is 3 cells. We numerically simulate the transmission spectrum of kink states for the topological resonator. As shown in Fig.~\ref{fig:2}(c), the resonator modes are observed in the whisper-gallery topological resonator. These modes have identical frequency separation, so-called the FSR. It is clear evidence that the topological resonator leads to an FSR of 450 GHz. A complete distinction is observed between topological bandgap and bulk states. For the region of bulk states, there is no resonator mode existing. Due to the excitations of valley kink states, topological optical frequency combs only appear at frequencies inside the topological bandgap. 

The Lorentzian fitting of the simulated resonant dip at the pump frequency of 370.67 THz reveals a total Q factor of $8.86\times10^4$ (Supplemental Material). The external loss $\kappa_{ex}$ and intrinsic loss $\kappa_{in}$ are calculated as $\kappa_{in}=\kappa_{ex}=1.35\times10^7 rad/s$. Correspondingly, the intrinsic quality factor $Q_{in}$ and external quality factor $Q_{ex}$ are calculated as $Q_{in}=Q_{ex}=1.772\times10^5$. Therefore, the topological resonator satisfies $\eta =1⁄2$, leading to the critical coupling. The threshold for parametric oscillation in the resonator is proportionable to $V_0⁄Q^2$, where $V_0$ is the effective volume of the pump. Hence, the generation of signal and idler sidebands from FWM is easy to manipulate in such high-Q VPC resonators.

Thanks to the intrinsic nonlinearity of the resonator, numerous sidebands with equidistant gaps are produced in the resonator with the injection of a CW pump laser at the angular frequency of $\omega_0$ (indexed by the number $\mu=0$). To visualize the stimulated FWM process of the topological resonator, we simulate the electric field profiles of the FWM process with the selected signal and idler frequencies ($\mu=1,-1$). Such numerical model of the FWM process in VPCs is performed in the software COMSOL Multiphysics (for details, see Supplemental Material). As depicted in Fig.~\ref{fig:3}, the kink states of the pump, signal, and idler are excited in the coupled waveguide-resonator device. A circularly polarized excitation is used to emulate kink states along the bus waveguide. Since the pump frequency is consistent with the resonator mode $\omega_0$, the injected pump is coupled to the topological resonator and excites the FWM process simultaneously. This chirality is also preserved for the cavity due to the same topology between the bus waveguide and the triangle cavity. There is no input for the kink state of the idler, thus, the generation of field profiles at the idler frequency gives clear evidence of the FWM processes~\cite{24}. The field profiles of topological resonator modes are concentrated at the interface, resulting in the high-efficiency generation of the FWM process. Remarkably, the resonator modes show robustness against sharp bends, confirming the topological nature of the QVH effect. Note that the direction of rotation in the topological ring is related to the injected pump mode (Supplemental Material).

\section{Topological Kerr soliton combs}

\begin{figure*}
\centering
\includegraphics[width=0.8\textwidth]{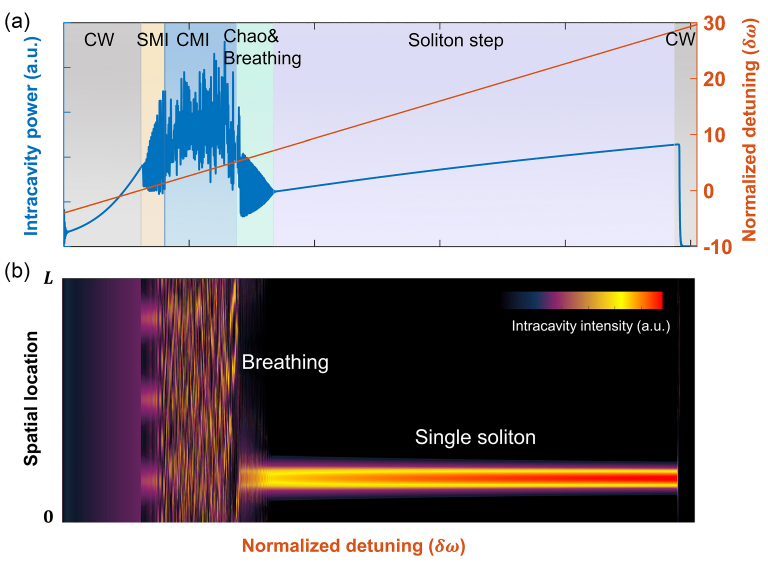}
\caption{Numerical simulated evolution of soliton formation in a topological resonator. (a) Effective intracavity energy evolution with the detuning from blue-detuned to red-detuned region. There appear three conspicuous processes including Turing rolls, chaotic states (breathing solitons), and single soliton states respectively. (b) Corresponding spatiotemporal evolution of soliton formation. }
    \label{fig:4}
\end{figure*}

DKS combs can be readily implemented in topological micro-resonators due to the intrinsic Kerr nonlinearity of nonlinear materials~\cite{17}. DKSs are temporal solutions to the balance between intrinsic dissipation and nonlinearity-induced parametric gain~\cite{1}. When the parametric gain of the cavity exceeds the decay rate, symmetrically spaced sidebands appear around the incident pump. Optical frequency combs can grow to DKSs with the tuning of the frequency of the pump laser. In general, the dynamics of optical frequency combs in the micro-resonator are described by Lugiato–Lefever equation (LLE)\cite{37,38}:

\begin{equation}
\begin{aligned}
    \frac{\partial}{\partial\tau}A=  
    &-\left(\frac{\kappa}{2}+i\delta\omega\right)A+i\pi\cdot FSR\cdot D_{2}\frac{\partial^{2}}{\partial T^{2}}A \\ 
     &+iL\cdot FSR\cdot\gamma|A|^{2}A+\sqrt{\frac{\kappa\eta P_{in}}{\hbar\omega}}, 
\end{aligned}
    \label{eq:3}
\end{equation}

where $A$ describes the evolution of the intracavity field, $\kappa$ is related to the linewidth of resonator modes, $\delta\omega$ is the detuning of the resonance frequency. $\eta=\kappa_{ex}/(\kappa_{ex}+\kappa_{in})$ is the coupling efficiency, where $\kappa_{in}$ and $\kappa_{ex}$ are the intrinsic loss and external loss respectively. 
With proper approximations of the LLE, the dynamics of DKSs can be numerically simulated (Supplemental Material). To access the reproducible soliton state, the pump is detuned from the blue ($\delta\omega<0$) to the red region ($\delta\omega>0$). In general, the evolution of optical frequency combs is recognized by several trajectories, including Turing rolls, chaotic states, breathing solitons, and eventually into the soliton states~\cite{1}. In the case of single soliton states, the comb spectrum is almost smooth and phase-locked~\cite{39}. Such single soliton states have been experimentally observed in a great deal of platforms~\cite{5,8,9,10}, therefore, we can conclude several tricks to excite single soliton states. First, solitons are always generated at the red tunning of resonance frequency due to the thermal effects of the cavity. Secondly, the combs of single soliton formation take parallel shapes of $\rm sech^2$ functions. Thirdly, in the time domain, multiple or single soliton state behaves with several spaced pulses or one ultrashort pulse upon a period of the round-trip time.

To numerically simulate the dynamic evolution of the topological resonator, we consider a whisper-gallery resonator as depicted in Fig.~\ref{fig:2}. The input pump frequency $\omega_0$ corresponds to the resonance frequency indexed by $\mu=0$, leading to the generation of the spaced sidebands in other resonator modes ($\mu\neq0$) via FWM processes. Since the pattern of the topological resonator is constructed by interfaces between $\rm VPC_1$ and $\rm VPC_2$, the optical frequency combs are generated at the interfaces, and they transport along a certain triangular path. However, when the incident pump is coincident with bulk bands, the output comb is barely visible. In the case of bulk modes, the incident pump is dissipated into the photonic crystals. Due to the topological protection of valley kink states, the backscattering around the sharp corners is negligible~\cite{40}. Therefore, the reflection of resonator modes at the sharp corners is not included in our theoretical model. 

\begin{figure*}
\centering
\includegraphics[width=0.7\textwidth]{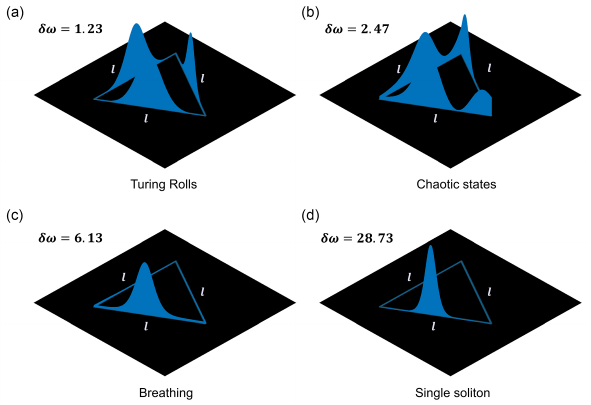}
\caption{Spatial intensity distribution of the (a) Turing rolls, (b) chaotic states, (c) breathing, and (d) single soliton state, respectively.}
    \label{fig:5}
\end{figure*}

In our simulation, the actual pump power is 2.2 W, and the refractive index of $\rm Si_{3}N_{4}$ is 2.0, with the nonlinear index $2.5\times10^{-19}\mathrm{m}^{2}\mathrm{W}^{-1}$. The FSR and dispersion $D_2$ are extracted from the simulated transmission (Fig.~\ref{fig:2}(c)), with calculated values of $FSR=450$ GHz and $ D_2=9.53$ GHz. The simulated total Q factor indicates a value of $8.86\times10^4$. The external loss and the total energy loss rate of the resonator are calculated as $\kappa_{ex}=1.35\times10^7$ rad/s and $\kappa=2.7\times10^7$ rad/s. In this case, the coupling efficiency of critical coupling is given by $\eta=1⁄2$. The length of the topological resonator is $L=3l$, where $l=180a$ is the side length of the triangular configuration. We assume the effective field cross-section area $A_{eff}=2.1\times10^{-14}~\mathrm{m}^{2}$, and the nonlinear coefficient of the topological resonator is given by $\gamma=\omega_{0}n_{2}/cA_{eff}$. We believe our topological DKSs can be experimentally accessed by present nanofabrication technology~\cite{39,41}.

The simulated dynamic evolution of topological DKSs as a function of detuning is illustrated in Fig.~\ref{fig:4}(a). And the corresponding spatiotemporal evolution of the DKS excitation process is shown in Fig.~\ref{fig:4}(b). Three conspicuous forms are identified during the evolution process, that is, Turing rolls, chaotic states (breathing solitons), and single soliton states, respectively. Fig.~\ref{fig:5} shows the spatial intensity distribution of the generated pulses with different detuning of the pump in the topological resonator, which can also be accessed from the spatiotemporal evolution (Supplemental Material). The pulses are circulated along the edge of the triangle topological resonator in a clockwise direction as time evolves.

In the early stages of evolution, the intensity build-up process leads to the increasing of intracavity power with the detuning from the blue side. The temporal pattern shows 3 equally spaced pulses within the triangular resonator, which is referred to as Turing rolls. Those stable pulses are attributed to the self-stabilization of the nonlinear evolution in the blue-detuned side of the resonance frequency. Thus, such a stage of modulation instability is referred to as stable modulation instability (SMI). We can observe that the cavity energy varies smoothly in the case of Turing rolls. 

\begin{figure*}
\centering
\includegraphics[width=0.7\textwidth]{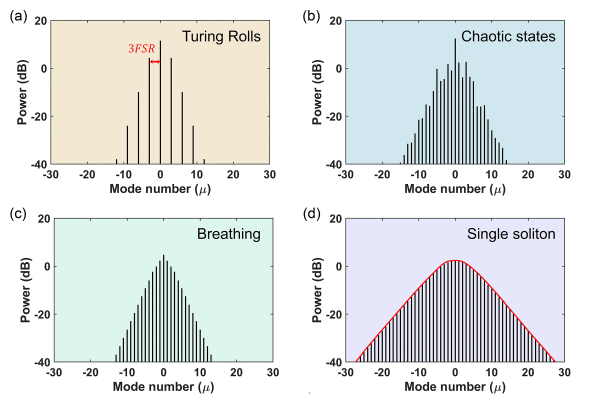}
\caption{Simulated optical frequency combs of (a) Turing rolls, (b) chaotic states, (c) breathing solitons, and (d) single soliton states.}
    \label{fig:6}
\end{figure*}

Those chaotic states are followed by the non-stationary behavior—breathing soliton states. Fig.~\ref{fig:5}(c) shows that there is only one pulse inside the topological resonator with the detuning $\delta\omega=6.13$, the amplitude and duration of the intracavity waveform are oscillated periodically, where the oscillation period is regarded as the breathing period. Correspondingly, the combs show periodical compression and stretching. Such instability occurs at the furcation of chaos and a stable DKS, which is referred to as Hopf bifurcation. 

When the pump is further detuned to the red side ($\delta\omega=28.73$), actual soliton formation can be accessed because the intracavity field is bistable. The intracavity power shows the formation of a “step” characteristic, this “step” is the identification of solitons. The “length” of the soliton step corresponds to the pump power. For the intracavity energy, the oscillation quickly evolves to a relatively stable situation. This bistable state is an elaborate consequence of the off-resonance pump detuning $\delta\omega$ and Kerr nonlinear shift of the resonator. 

As shown in Fig.~\ref{fig:4}(b), the corresponding spatiotemporal formation of single soliton states also exhibits intriguing results. A quick comparison with the spatiotemporal intensity distribution of Turing rolls and chaotic states reveals that there exists only one ultrashort pulse upon a period of the round-trip time for a single soliton state. Once a single soliton state is accessed, the thermal nonlinearity of solitons makes the laser-cavity detuning self-stabilized. In this scenario, single soliton states can be stable for several hours. Note that multiple solitons may exist in the cavity with a stochastic separation number. Proceeding further, the comb degenerates to a continuous component due to the separation of the pump and resonance frequency.

The formations of corresponding optical frequency combs are shown in Fig.~\ref{fig:6}. It is noted that generated Kerr frequency combs are composed of a CW pump component and relatively weak bilateral combs~\cite{6}. When the nonlinear gain overcomes the cavity losses, the developed primary comb so-called Turing rolls is generated. As seen in Fig.~\ref{fig:6}(a), the comb amplitudes of Turing rolls have distinctly contrast. 

When the pump is further detuned to the red side, the increasing intracavity power initiates FWM processes, which leads to secondary sidebands and subcombs. The output comb of chaotic states is depicted in Fig.~\ref{fig:6}(b), Random variations of combs indicate that there is no stationary solution existing, leading to the incoherence of the waveform.

Following chaotic states, the generation of breathing solitons means the arrival of a low-detuning boundary of the soliton. As depicted in Fig.~\ref{fig:6}(c), the comb of breathing solitons takes the form of a triangular feature. Note that the amplitudes of the comb oscillate with the breathing period, and the comb returns to its initial state after one period.

The self-sustaining wavepackets of the single soliton state indicate that comb lines are phase-locked with each other (Fig.~\ref{fig:6}(d)). More importantly, the shape of the spectral output takes the form of $\rm sech^2$ functions. This behavior of a single soliton state in the topological resonator is analogous to that observed in $\rm Si_{3}N_{4}$ single-ring resonators where exist single solitons. 

We have shown many resemblances between topological resonators and single-ring resonators, however, for the DKS and single soliton combs produced in our design, thanks to the topological nature of VPCs, they are inherited to be robust against sharp bends and certain disorders. In the experimental realization, the strong thermal effect becomes an intractable element to produce stable solitons in the red-detuned region. To overcome this, heaters can be used to control the temperature of VPC chips~\cite{39,41}. Based on the state-of-the-art nanofabrication technology, our topological DKS combs design could be implemented experimentally.

\section{Conclusion}
In this work, we have demonstrated a theoretical scheme exhibiting topological optical frequency combs and DKSs combs. The triangle resonator is composed of VPCs with different topologies and can excite topological valley kink resonator modes propagating along the interface. We numerically simulate the nonlinear dynamic evolution of the topological resonator with the injection of the pump laser. The result reveals that the single soliton states can be produced in the resonator, and they are born to be topological. This topological nature endows the DKS combs’ robustness against sharp bends and disorders. Our topological frequency combs could be readily accessed in on-chip nanofabricated photonics.

\begin{acknowledgments}
This work is supported by the Key-Area Research and Development Program of Guangdong Province (2018B030325002), the National Natural Science Foundation of China (62075129, 61975119), the SJTU Pinghu Institute of Intelligent Optoelectronics (2022SPIOE204), and the Science and Technology on Metrology and Calibration Laboratory (JLJK2022001B002).
\end{acknowledgments}


\begin{thebibliography}{41}%
\makeatletter
\providecommand \@ifxundefined [1]{%
 \@ifx{#1\undefined}
}%
\providecommand \@ifnum [1]{%
 \ifnum #1\expandafter \@firstoftwo
 \else \expandafter \@secondoftwo
 \fi
}%
\providecommand \@ifx [1]{%
 \ifx #1\expandafter \@firstoftwo
 \else \expandafter \@secondoftwo
 \fi
}%
\providecommand \natexlab [1]{#1}%
\providecommand \enquote  [1]{``#1''}%
\providecommand \bibnamefont  [1]{#1}%
\providecommand \bibfnamefont [1]{#1}%
\providecommand \citenamefont [1]{#1}%
\providecommand \href@noop [0]{\@secondoftwo}%
\providecommand \href [0]{\begingroup \@sanitize@url \@href}%
\providecommand \@href[1]{\@@startlink{#1}\@@href}%
\providecommand \@@href[1]{\endgroup#1\@@endlink}%
\providecommand \@sanitize@url [0]{\catcode `\\12\catcode `\$12\catcode
  `\&12\catcode `\#12\catcode `\^12\catcode `\_12\catcode `\%12\relax}%
\providecommand \@@startlink[1]{}%
\providecommand \@@endlink[0]{}%
\providecommand \url  [0]{\begingroup\@sanitize@url \@url }%
\providecommand \@url [1]{\endgroup\@href {#1}{\urlprefix }}%
\providecommand \urlprefix  [0]{URL }%
\providecommand \Eprint [0]{\href }%
\providecommand \doibase [0]{https://doi.org/}%
\providecommand \selectlanguage [0]{\@gobble}%
\providecommand \bibinfo  [0]{\@secondoftwo}%
\providecommand \bibfield  [0]{\@secondoftwo}%
\providecommand \translation [1]{[#1]}%
\providecommand \BibitemOpen [0]{}%
\providecommand \bibitemStop [0]{}%
\providecommand \bibitemNoStop [0]{.\EOS\space}%
\providecommand \EOS [0]{\spacefactor3000\relax}%
\providecommand \BibitemShut  [1]{\csname bibitem#1\endcsname}%
\let\auto@bib@innerbib\@empty
\bibitem [{\citenamefont {Kippenberg}\ \emph {et~al.}(2018)\citenamefont
  {Kippenberg}, \citenamefont {Gaeta}, \citenamefont {Lipson},\ and\
  \citenamefont {Gorodetsky}}]{1}%
  \BibitemOpen
  \bibfield  {author} {\bibinfo {author} {\bibfnamefont {T.~J.}\ \bibnamefont
  {Kippenberg}}, \bibinfo {author} {\bibfnamefont {A.~L.}\ \bibnamefont
  {Gaeta}}, \bibinfo {author} {\bibfnamefont {M.}~\bibnamefont {Lipson}},\ and\
  \bibinfo {author} {\bibfnamefont {M.~L.}\ \bibnamefont {Gorodetsky}},\
  }\bibfield  {title} {\bibinfo {title} {Dissipative kerr solitons in optical
  microresonators},\ }\href@noop {} {\bibfield  {journal} {\bibinfo  {journal}
  {Science}\ }\textbf {\bibinfo {volume} {361}},\ \bibinfo {pages} {eaan8083}
  (\bibinfo {year} {2018})}\BibitemShut {NoStop}%
\bibitem [{\citenamefont {Yang}\ \emph {et~al.}(2017)\citenamefont {Yang},
  \citenamefont {Yi}, \citenamefont {Yang},\ and\ \citenamefont {Vahala}}]{2}%
  \BibitemOpen
  \bibfield  {author} {\bibinfo {author} {\bibfnamefont {Q.-F.}\ \bibnamefont
  {Yang}}, \bibinfo {author} {\bibfnamefont {X.}~\bibnamefont {Yi}}, \bibinfo
  {author} {\bibfnamefont {K.~Y.}\ \bibnamefont {Yang}},\ and\ \bibinfo
  {author} {\bibfnamefont {K.}~\bibnamefont {Vahala}},\ }\bibfield  {title}
  {\bibinfo {title} {Stokes solitons in optical microcavities},\ }\href@noop {}
  {\bibfield  {journal} {\bibinfo  {journal} {Nature Physics}\ }\textbf
  {\bibinfo {volume} {13}},\ \bibinfo {pages} {53} (\bibinfo {year}
  {2017})}\BibitemShut {NoStop}%
\bibitem [{\citenamefont {Bao}\ \emph {et~al.}(2016)\citenamefont {Bao},
  \citenamefont {Jaramillo-Villegas}, \citenamefont {Xuan}, \citenamefont
  {Leaird}, \citenamefont {Qi},\ and\ \citenamefont {Weiner}}]{3}%
  \BibitemOpen
  \bibfield  {author} {\bibinfo {author} {\bibfnamefont {C.}~\bibnamefont
  {Bao}}, \bibinfo {author} {\bibfnamefont {J.~A.}\ \bibnamefont
  {Jaramillo-Villegas}}, \bibinfo {author} {\bibfnamefont {Y.}~\bibnamefont
  {Xuan}}, \bibinfo {author} {\bibfnamefont {D.~E.}\ \bibnamefont {Leaird}},
  \bibinfo {author} {\bibfnamefont {M.}~\bibnamefont {Qi}},\ and\ \bibinfo
  {author} {\bibfnamefont {A.~M.}\ \bibnamefont {Weiner}},\ }\bibfield  {title}
  {\bibinfo {title} {Observation of fermi-pasta-ulam recurrence induced by
  breather solitons in an optical microresonator},\ }\href@noop {} {\bibfield
  {journal} {\bibinfo  {journal} {Physical Review Letters}\ }\textbf {\bibinfo
  {volume} {117}},\ \bibinfo {pages} {163901} (\bibinfo {year}
  {2016})}\BibitemShut {NoStop}%
\bibitem [{\citenamefont {Cole}\ \emph {et~al.}(2017)\citenamefont {Cole},
  \citenamefont {Lamb}, \citenamefont {Del’Haye}, \citenamefont {Diddams},\
  and\ \citenamefont {Papp}}]{4}%
  \BibitemOpen
  \bibfield  {author} {\bibinfo {author} {\bibfnamefont {D.~C.}\ \bibnamefont
  {Cole}}, \bibinfo {author} {\bibfnamefont {E.~S.}\ \bibnamefont {Lamb}},
  \bibinfo {author} {\bibfnamefont {P.}~\bibnamefont {Del’Haye}}, \bibinfo
  {author} {\bibfnamefont {S.~A.}\ \bibnamefont {Diddams}},\ and\ \bibinfo
  {author} {\bibfnamefont {S.~B.}\ \bibnamefont {Papp}},\ }\bibfield  {title}
  {\bibinfo {title} {Soliton crystals in kerr resonators},\ }\href@noop {}
  {\bibfield  {journal} {\bibinfo  {journal} {Nature Photonics}\ }\textbf
  {\bibinfo {volume} {11}},\ \bibinfo {pages} {671} (\bibinfo {year}
  {2017})}\BibitemShut {NoStop}%
\bibitem [{\citenamefont {Liang}\ \emph {et~al.}(2015)\citenamefont {Liang},
  \citenamefont {Eliyahu}, \citenamefont {Ilchenko}, \citenamefont
  {Savchenkov}, \citenamefont {Matsko}, \citenamefont {Seidel},\ and\
  \citenamefont {Maleki}}]{5}%
  \BibitemOpen
  \bibfield  {author} {\bibinfo {author} {\bibfnamefont {W.}~\bibnamefont
  {Liang}}, \bibinfo {author} {\bibfnamefont {D.}~\bibnamefont {Eliyahu}},
  \bibinfo {author} {\bibfnamefont {V.~S.}\ \bibnamefont {Ilchenko}}, \bibinfo
  {author} {\bibfnamefont {A.~A.}\ \bibnamefont {Savchenkov}}, \bibinfo
  {author} {\bibfnamefont {A.~B.}\ \bibnamefont {Matsko}}, \bibinfo {author}
  {\bibfnamefont {D.}~\bibnamefont {Seidel}},\ and\ \bibinfo {author}
  {\bibfnamefont {L.}~\bibnamefont {Maleki}},\ }\bibfield  {title} {\bibinfo
  {title} {High spectral purity kerr frequency comb radio frequency photonic
  oscillator},\ }\href@noop {} {\bibfield  {journal} {\bibinfo  {journal}
  {Nature Communications}\ }\textbf {\bibinfo {volume} {6}},\ \bibinfo {pages}
  {7957} (\bibinfo {year} {2015})}\BibitemShut {NoStop}%
\bibitem [{\citenamefont {Herr}\ \emph {et~al.}(2014)\citenamefont {Herr},
  \citenamefont {Brasch}, \citenamefont {Jost}, \citenamefont {Wang},
  \citenamefont {Kondratiev}, \citenamefont {Gorodetsky},\ and\ \citenamefont
  {Kippenberg}}]{6}%
  \BibitemOpen
  \bibfield  {author} {\bibinfo {author} {\bibfnamefont {T.}~\bibnamefont
  {Herr}}, \bibinfo {author} {\bibfnamefont {V.}~\bibnamefont {Brasch}},
  \bibinfo {author} {\bibfnamefont {J.~D.}\ \bibnamefont {Jost}}, \bibinfo
  {author} {\bibfnamefont {C.~Y.}\ \bibnamefont {Wang}}, \bibinfo {author}
  {\bibfnamefont {N.~M.}\ \bibnamefont {Kondratiev}}, \bibinfo {author}
  {\bibfnamefont {M.~L.}\ \bibnamefont {Gorodetsky}},\ and\ \bibinfo {author}
  {\bibfnamefont {T.~J.}\ \bibnamefont {Kippenberg}},\ }\bibfield  {title}
  {\bibinfo {title} {Temporal solitons in optical microresonators},\
  }\href@noop {} {\bibfield  {journal} {\bibinfo  {journal} {Nature Photonics}\
  }\textbf {\bibinfo {volume} {8}},\ \bibinfo {pages} {145} (\bibinfo {year}
  {2014})}\BibitemShut {NoStop}%
\bibitem [{\citenamefont {Yi}\ \emph {et~al.}(2015)\citenamefont {Yi},
  \citenamefont {Yang}, \citenamefont {Yang}, \citenamefont {Suh},\ and\
  \citenamefont {Vahala}}]{7}%
  \BibitemOpen
  \bibfield  {author} {\bibinfo {author} {\bibfnamefont {X.}~\bibnamefont
  {Yi}}, \bibinfo {author} {\bibfnamefont {Q.-F.}\ \bibnamefont {Yang}},
  \bibinfo {author} {\bibfnamefont {K.~Y.}\ \bibnamefont {Yang}}, \bibinfo
  {author} {\bibfnamefont {M.-G.}\ \bibnamefont {Suh}},\ and\ \bibinfo {author}
  {\bibfnamefont {K.}~\bibnamefont {Vahala}},\ }\bibfield  {title} {\bibinfo
  {title} {Soliton frequency comb at microwave rates in a high-q silica
  microresonator},\ }\href@noop {} {\bibfield  {journal} {\bibinfo  {journal}
  {Optica}\ }\textbf {\bibinfo {volume} {2}},\ \bibinfo {pages} {1078}
  (\bibinfo {year} {2015})}\BibitemShut {NoStop}%
\bibitem [{\citenamefont {Brasch}\ \emph {et~al.}(2016)\citenamefont {Brasch},
  \citenamefont {Geiselmann}, \citenamefont {Herr}, \citenamefont {Lihachev},
  \citenamefont {Pfeiffer}, \citenamefont {Gorodetsky},\ and\ \citenamefont
  {Kippenberg}}]{8}%
  \BibitemOpen
  \bibfield  {author} {\bibinfo {author} {\bibfnamefont {V.}~\bibnamefont
  {Brasch}}, \bibinfo {author} {\bibfnamefont {M.}~\bibnamefont {Geiselmann}},
  \bibinfo {author} {\bibfnamefont {T.}~\bibnamefont {Herr}}, \bibinfo {author}
  {\bibfnamefont {G.}~\bibnamefont {Lihachev}}, \bibinfo {author}
  {\bibfnamefont {M.~H.}\ \bibnamefont {Pfeiffer}}, \bibinfo {author}
  {\bibfnamefont {M.~L.}\ \bibnamefont {Gorodetsky}},\ and\ \bibinfo {author}
  {\bibfnamefont {T.~J.}\ \bibnamefont {Kippenberg}},\ }\bibfield  {title}
  {\bibinfo {title} {Photonic chip--based optical frequency comb using soliton
  cherenkov radiation},\ }\href@noop {} {\bibfield  {journal} {\bibinfo
  {journal} {Science}\ }\textbf {\bibinfo {volume} {351}},\ \bibinfo {pages}
  {357} (\bibinfo {year} {2016})}\BibitemShut {NoStop}%
\bibitem [{\citenamefont {Zhou}\ \emph {et~al.}(2022)\citenamefont {Zhou},
  \citenamefont {Shen}, \citenamefont {Xi}, \citenamefont {Huang},\ and\
  \citenamefont {He}}]{9}%
  \BibitemOpen
  \bibfield  {author} {\bibinfo {author} {\bibfnamefont {L.}~\bibnamefont
  {Zhou}}, \bibinfo {author} {\bibfnamefont {Y.}~\bibnamefont {Shen}}, \bibinfo
  {author} {\bibfnamefont {C.}~\bibnamefont {Xi}}, \bibinfo {author}
  {\bibfnamefont {X.}~\bibnamefont {Huang}},\ and\ \bibinfo {author}
  {\bibfnamefont {G.}~\bibnamefont {He}},\ }\bibfield  {title} {\bibinfo
  {title} {Computer-controlled microresonator soliton comb system automating
  soliton generation and expanding excursion bandwidth},\ }\href@noop {}
  {\bibfield  {journal} {\bibinfo  {journal} {Optics Continuum}\ }\textbf
  {\bibinfo {volume} {1}},\ \bibinfo {pages} {161} (\bibinfo {year}
  {2022})}\BibitemShut {NoStop}%
\bibitem [{\citenamefont {Joshi}\ \emph {et~al.}(2016)\citenamefont {Joshi},
  \citenamefont {Jang}, \citenamefont {Luke}, \citenamefont {Ji}, \citenamefont
  {Miller}, \citenamefont {Klenner}, \citenamefont {Okawachi}, \citenamefont
  {Lipson},\ and\ \citenamefont {Gaeta}}]{10}%
  \BibitemOpen
  \bibfield  {author} {\bibinfo {author} {\bibfnamefont {C.}~\bibnamefont
  {Joshi}}, \bibinfo {author} {\bibfnamefont {J.~K.}\ \bibnamefont {Jang}},
  \bibinfo {author} {\bibfnamefont {K.}~\bibnamefont {Luke}}, \bibinfo {author}
  {\bibfnamefont {X.}~\bibnamefont {Ji}}, \bibinfo {author} {\bibfnamefont
  {S.~A.}\ \bibnamefont {Miller}}, \bibinfo {author} {\bibfnamefont
  {A.}~\bibnamefont {Klenner}}, \bibinfo {author} {\bibfnamefont
  {Y.}~\bibnamefont {Okawachi}}, \bibinfo {author} {\bibfnamefont
  {M.}~\bibnamefont {Lipson}},\ and\ \bibinfo {author} {\bibfnamefont {A.~L.}\
  \bibnamefont {Gaeta}},\ }\bibfield  {title} {\bibinfo {title} {Thermally
  controlled comb generation and soliton modelocking in microresonators},\
  }\href@noop {} {\bibfield  {journal} {\bibinfo  {journal} {Optics Letters}\
  }\textbf {\bibinfo {volume} {41}},\ \bibinfo {pages} {2565} (\bibinfo {year}
  {2016})}\BibitemShut {NoStop}%
\bibitem [{\citenamefont {He}\ \emph {et~al.}(2019{\natexlab{a}})\citenamefont
  {He}, \citenamefont {Yang}, \citenamefont {Ling}, \citenamefont {Luo},
  \citenamefont {Liang}, \citenamefont {Li}, \citenamefont {Shen},
  \citenamefont {Wang}, \citenamefont {Vahala},\ and\ \citenamefont
  {Lin}}]{11}%
  \BibitemOpen
  \bibfield  {author} {\bibinfo {author} {\bibfnamefont {Y.}~\bibnamefont
  {He}}, \bibinfo {author} {\bibfnamefont {Q.-F.}\ \bibnamefont {Yang}},
  \bibinfo {author} {\bibfnamefont {J.}~\bibnamefont {Ling}}, \bibinfo {author}
  {\bibfnamefont {R.}~\bibnamefont {Luo}}, \bibinfo {author} {\bibfnamefont
  {H.}~\bibnamefont {Liang}}, \bibinfo {author} {\bibfnamefont
  {M.}~\bibnamefont {Li}}, \bibinfo {author} {\bibfnamefont {B.}~\bibnamefont
  {Shen}}, \bibinfo {author} {\bibfnamefont {H.}~\bibnamefont {Wang}}, \bibinfo
  {author} {\bibfnamefont {K.}~\bibnamefont {Vahala}},\ and\ \bibinfo {author}
  {\bibfnamefont {Q.}~\bibnamefont {Lin}},\ }\bibfield  {title} {\bibinfo
  {title} {Self-starting bi-chromatic linbo 3 soliton microcomb},\ }\href@noop
  {} {\bibfield  {journal} {\bibinfo  {journal} {Optica}\ }\textbf {\bibinfo
  {volume} {6}},\ \bibinfo {pages} {1138} (\bibinfo {year}
  {2019}{\natexlab{a}})}\BibitemShut {NoStop}%
\bibitem [{\citenamefont {Gong}\ \emph {et~al.}(2019)\citenamefont {Gong},
  \citenamefont {Liu}, \citenamefont {Xu}, \citenamefont {Xu}, \citenamefont
  {Surya}, \citenamefont {Lu}, \citenamefont {Bruch}, \citenamefont {Zou},\
  and\ \citenamefont {Tang}}]{12}%
  \BibitemOpen
  \bibfield  {author} {\bibinfo {author} {\bibfnamefont {Z.}~\bibnamefont
  {Gong}}, \bibinfo {author} {\bibfnamefont {X.}~\bibnamefont {Liu}}, \bibinfo
  {author} {\bibfnamefont {Y.}~\bibnamefont {Xu}}, \bibinfo {author}
  {\bibfnamefont {M.}~\bibnamefont {Xu}}, \bibinfo {author} {\bibfnamefont
  {J.~B.}\ \bibnamefont {Surya}}, \bibinfo {author} {\bibfnamefont
  {J.}~\bibnamefont {Lu}}, \bibinfo {author} {\bibfnamefont {A.}~\bibnamefont
  {Bruch}}, \bibinfo {author} {\bibfnamefont {C.}~\bibnamefont {Zou}},\ and\
  \bibinfo {author} {\bibfnamefont {H.~X.}\ \bibnamefont {Tang}},\ }\bibfield
  {title} {\bibinfo {title} {Soliton microcomb generation at 2 $\mu$m in z-cut
  lithium niobate microring resonators},\ }\href@noop {} {\bibfield  {journal}
  {\bibinfo  {journal} {Optics Letters}\ }\textbf {\bibinfo {volume} {44}},\
  \bibinfo {pages} {3182} (\bibinfo {year} {2019})}\BibitemShut {NoStop}%
\bibitem [{\citenamefont {Pu}\ \emph {et~al.}(2016)\citenamefont {Pu},
  \citenamefont {Ottaviano}, \citenamefont {Semenova},\ and\ \citenamefont
  {Yvind}}]{13}%
  \BibitemOpen
  \bibfield  {author} {\bibinfo {author} {\bibfnamefont {M.}~\bibnamefont
  {Pu}}, \bibinfo {author} {\bibfnamefont {L.}~\bibnamefont {Ottaviano}},
  \bibinfo {author} {\bibfnamefont {E.}~\bibnamefont {Semenova}},\ and\
  \bibinfo {author} {\bibfnamefont {K.}~\bibnamefont {Yvind}},\ }\bibfield
  {title} {\bibinfo {title} {Efficient frequency comb generation in
  algaas-on-insulator},\ }\href@noop {} {\bibfield  {journal} {\bibinfo
  {journal} {Optica}\ }\textbf {\bibinfo {volume} {3}},\ \bibinfo {pages} {823}
  (\bibinfo {year} {2016})}\BibitemShut {NoStop}%
\bibitem [{\citenamefont {Smirnova}\ \emph
  {et~al.}(2019{\natexlab{a}})\citenamefont {Smirnova}, \citenamefont {Kruk},
  \citenamefont {Leykam}, \citenamefont {Melik-Gaykazyan}, \citenamefont
  {Choi},\ and\ \citenamefont {Kivshar}}]{14}%
  \BibitemOpen
  \bibfield  {author} {\bibinfo {author} {\bibfnamefont {D.}~\bibnamefont
  {Smirnova}}, \bibinfo {author} {\bibfnamefont {S.}~\bibnamefont {Kruk}},
  \bibinfo {author} {\bibfnamefont {D.}~\bibnamefont {Leykam}}, \bibinfo
  {author} {\bibfnamefont {E.}~\bibnamefont {Melik-Gaykazyan}}, \bibinfo
  {author} {\bibfnamefont {D.-Y.}\ \bibnamefont {Choi}},\ and\ \bibinfo
  {author} {\bibfnamefont {Y.}~\bibnamefont {Kivshar}},\ }\bibfield  {title}
  {\bibinfo {title} {Third-harmonic generation in photonic topological
  metasurfaces},\ }\href@noop {} {\bibfield  {journal} {\bibinfo  {journal}
  {Physical Review Letters}\ }\textbf {\bibinfo {volume} {123}},\ \bibinfo
  {pages} {103901} (\bibinfo {year} {2019}{\natexlab{a}})}\BibitemShut
  {NoStop}%
\bibitem [{\citenamefont {Kruk}\ \emph {et~al.}(2021)\citenamefont {Kruk},
  \citenamefont {Gao}, \citenamefont {Choi}, \citenamefont {Zentgraf},
  \citenamefont {Zhang},\ and\ \citenamefont {Kivshar}}]{15}%
  \BibitemOpen
  \bibfield  {author} {\bibinfo {author} {\bibfnamefont {S.~S.}\ \bibnamefont
  {Kruk}}, \bibinfo {author} {\bibfnamefont {W.}~\bibnamefont {Gao}}, \bibinfo
  {author} {\bibfnamefont {D.-Y.}\ \bibnamefont {Choi}}, \bibinfo {author}
  {\bibfnamefont {T.}~\bibnamefont {Zentgraf}}, \bibinfo {author}
  {\bibfnamefont {S.}~\bibnamefont {Zhang}},\ and\ \bibinfo {author}
  {\bibfnamefont {Y.}~\bibnamefont {Kivshar}},\ }\bibfield  {title} {\bibinfo
  {title} {Nonlinear imaging of nanoscale topological corner states},\
  }\href@noop {} {\bibfield  {journal} {\bibinfo  {journal} {Nano Letters}\
  }\textbf {\bibinfo {volume} {21}},\ \bibinfo {pages} {4592} (\bibinfo {year}
  {2021})}\BibitemShut {NoStop}%
\bibitem [{\citenamefont {You}\ \emph {et~al.}(2020)\citenamefont {You},
  \citenamefont {Lan},\ and\ \citenamefont {Panoiu}}]{16}%
  \BibitemOpen
  \bibfield  {author} {\bibinfo {author} {\bibfnamefont {J.~W.}\ \bibnamefont
  {You}}, \bibinfo {author} {\bibfnamefont {Z.}~\bibnamefont {Lan}},\ and\
  \bibinfo {author} {\bibfnamefont {N.~C.}\ \bibnamefont {Panoiu}},\ }\bibfield
   {title} {\bibinfo {title} {Four-wave mixing of topological edge plasmons in
  graphene metasurfaces},\ }\href@noop {} {\bibfield  {journal} {\bibinfo
  {journal} {Science Advances}\ }\textbf {\bibinfo {volume} {6}},\ \bibinfo
  {pages} {eaaz3910} (\bibinfo {year} {2020})}\BibitemShut {NoStop}%
\bibitem [{\citenamefont {Mittal}\ \emph
  {et~al.}(2021{\natexlab{a}})\citenamefont {Mittal}, \citenamefont {Moille},
  \citenamefont {Srinivasan}, \citenamefont {Chembo},\ and\ \citenamefont
  {Hafezi}}]{17}%
  \BibitemOpen
  \bibfield  {author} {\bibinfo {author} {\bibfnamefont {S.}~\bibnamefont
  {Mittal}}, \bibinfo {author} {\bibfnamefont {G.}~\bibnamefont {Moille}},
  \bibinfo {author} {\bibfnamefont {K.}~\bibnamefont {Srinivasan}}, \bibinfo
  {author} {\bibfnamefont {Y.~K.}\ \bibnamefont {Chembo}},\ and\ \bibinfo
  {author} {\bibfnamefont {M.}~\bibnamefont {Hafezi}},\ }\bibfield  {title}
  {\bibinfo {title} {Topological frequency combs and nested temporal
  solitons},\ }\href@noop {} {\bibfield  {journal} {\bibinfo  {journal} {Nature
  Physics}\ }\textbf {\bibinfo {volume} {17}},\ \bibinfo {pages} {1169}
  (\bibinfo {year} {2021}{\natexlab{a}})}\BibitemShut {NoStop}%
\bibitem [{\citenamefont {Kartashov}\ and\ \citenamefont
  {Skryabin}(2017)}]{18}%
  \BibitemOpen
  \bibfield  {author} {\bibinfo {author} {\bibfnamefont {Y.~V.}\ \bibnamefont
  {Kartashov}}\ and\ \bibinfo {author} {\bibfnamefont {D.~V.}\ \bibnamefont
  {Skryabin}},\ }\bibfield  {title} {\bibinfo {title} {Bistable topological
  insulator with exciton-polaritons},\ }\href@noop {} {\bibfield  {journal}
  {\bibinfo  {journal} {Physical Review Letters}\ }\textbf {\bibinfo {volume}
  {119}},\ \bibinfo {pages} {253904} (\bibinfo {year} {2017})}\BibitemShut
  {NoStop}%
\bibitem [{\citenamefont {Banerjee}\ \emph {et~al.}(2020)\citenamefont
  {Banerjee}, \citenamefont {Mandal},\ and\ \citenamefont {Liew}}]{19}%
  \BibitemOpen
  \bibfield  {author} {\bibinfo {author} {\bibfnamefont {R.}~\bibnamefont
  {Banerjee}}, \bibinfo {author} {\bibfnamefont {S.}~\bibnamefont {Mandal}},\
  and\ \bibinfo {author} {\bibfnamefont {T.}~\bibnamefont {Liew}},\ }\bibfield
  {title} {\bibinfo {title} {Coupling between exciton-polariton corner modes
  through edge states},\ }\href@noop {} {\bibfield  {journal} {\bibinfo
  {journal} {Physical Review Letters}\ }\textbf {\bibinfo {volume} {124}},\
  \bibinfo {pages} {063901} (\bibinfo {year} {2020})}\BibitemShut {NoStop}%
\bibitem [{\citenamefont {Smirnova}\ \emph
  {et~al.}(2019{\natexlab{b}})\citenamefont {Smirnova}, \citenamefont
  {Smirnov}, \citenamefont {Leykam},\ and\ \citenamefont {Kivshar}}]{20}%
  \BibitemOpen
  \bibfield  {author} {\bibinfo {author} {\bibfnamefont {D.~A.}\ \bibnamefont
  {Smirnova}}, \bibinfo {author} {\bibfnamefont {L.~A.}\ \bibnamefont
  {Smirnov}}, \bibinfo {author} {\bibfnamefont {D.}~\bibnamefont {Leykam}},\
  and\ \bibinfo {author} {\bibfnamefont {Y.~S.}\ \bibnamefont {Kivshar}},\
  }\bibfield  {title} {\bibinfo {title} {Topological edge states and gap
  solitons in the nonlinear dirac model},\ }\href@noop {} {\bibfield  {journal}
  {\bibinfo  {journal} {Laser \& Photonics Reviews}\ }\textbf {\bibinfo
  {volume} {13}},\ \bibinfo {pages} {1900223} (\bibinfo {year}
  {2019}{\natexlab{b}})}\BibitemShut {NoStop}%
\bibitem [{\citenamefont {Pernet}\ \emph {et~al.}(2022)\citenamefont {Pernet},
  \citenamefont {St-Jean}, \citenamefont {Solnyshkov}, \citenamefont
  {Malpuech}, \citenamefont {Carlon~Zambon}, \citenamefont {Fontaine},
  \citenamefont {Real}, \citenamefont {Jamadi}, \citenamefont {Lemaitre},
  \citenamefont {Morassi} \emph {et~al.}}]{21}%
  \BibitemOpen
  \bibfield  {author} {\bibinfo {author} {\bibfnamefont {N.}~\bibnamefont
  {Pernet}}, \bibinfo {author} {\bibfnamefont {P.}~\bibnamefont {St-Jean}},
  \bibinfo {author} {\bibfnamefont {D.~D.}\ \bibnamefont {Solnyshkov}},
  \bibinfo {author} {\bibfnamefont {G.}~\bibnamefont {Malpuech}}, \bibinfo
  {author} {\bibfnamefont {N.}~\bibnamefont {Carlon~Zambon}}, \bibinfo {author}
  {\bibfnamefont {Q.}~\bibnamefont {Fontaine}}, \bibinfo {author}
  {\bibfnamefont {B.}~\bibnamefont {Real}}, \bibinfo {author} {\bibfnamefont
  {O.}~\bibnamefont {Jamadi}}, \bibinfo {author} {\bibfnamefont
  {A.}~\bibnamefont {Lemaitre}}, \bibinfo {author} {\bibfnamefont
  {M.}~\bibnamefont {Morassi}}, \emph {et~al.},\ }\bibfield  {title} {\bibinfo
  {title} {Gap solitons in a one-dimensional driven-dissipative topological
  lattice},\ }\href@noop {} {\bibfield  {journal} {\bibinfo  {journal} {Nature
  Physics}\ }\textbf {\bibinfo {volume} {18}},\ \bibinfo {pages} {678}
  (\bibinfo {year} {2022})}\BibitemShut {NoStop}%
\bibitem [{\citenamefont {Mittal}\ \emph {et~al.}(2018)\citenamefont {Mittal},
  \citenamefont {Goldschmidt},\ and\ \citenamefont {Hafezi}}]{22}%
  \BibitemOpen
  \bibfield  {author} {\bibinfo {author} {\bibfnamefont {S.}~\bibnamefont
  {Mittal}}, \bibinfo {author} {\bibfnamefont {E.~A.}\ \bibnamefont
  {Goldschmidt}},\ and\ \bibinfo {author} {\bibfnamefont {M.}~\bibnamefont
  {Hafezi}},\ }\bibfield  {title} {\bibinfo {title} {A topological source of
  quantum light},\ }\href@noop {} {\bibfield  {journal} {\bibinfo  {journal}
  {Nature}\ }\textbf {\bibinfo {volume} {561}},\ \bibinfo {pages} {502}
  (\bibinfo {year} {2018})}\BibitemShut {NoStop}%
\bibitem [{\citenamefont {Chen}\ \emph {et~al.}(2021)\citenamefont {Chen},
  \citenamefont {He}, \citenamefont {Cheng}, \citenamefont {Qiu}, \citenamefont
  {Feng}, \citenamefont {Zhang}, \citenamefont {Dai}, \citenamefont {Guo},
  \citenamefont {Dong},\ and\ \citenamefont {Ren}}]{23}%
  \BibitemOpen
  \bibfield  {author} {\bibinfo {author} {\bibfnamefont {Y.}~\bibnamefont
  {Chen}}, \bibinfo {author} {\bibfnamefont {X.-T.}\ \bibnamefont {He}},
  \bibinfo {author} {\bibfnamefont {Y.-J.}\ \bibnamefont {Cheng}}, \bibinfo
  {author} {\bibfnamefont {H.-Y.}\ \bibnamefont {Qiu}}, \bibinfo {author}
  {\bibfnamefont {L.-T.}\ \bibnamefont {Feng}}, \bibinfo {author}
  {\bibfnamefont {M.}~\bibnamefont {Zhang}}, \bibinfo {author} {\bibfnamefont
  {D.-X.}\ \bibnamefont {Dai}}, \bibinfo {author} {\bibfnamefont {G.-C.}\
  \bibnamefont {Guo}}, \bibinfo {author} {\bibfnamefont {J.-W.}\ \bibnamefont
  {Dong}},\ and\ \bibinfo {author} {\bibfnamefont {X.-F.}\ \bibnamefont
  {Ren}},\ }\bibfield  {title} {\bibinfo {title} {Topologically protected
  valley-dependent quantum photonic circuits},\ }\href@noop {} {\bibfield
  {journal} {\bibinfo  {journal} {Physical Review Letters}\ }\textbf {\bibinfo
  {volume} {126}},\ \bibinfo {pages} {230503} (\bibinfo {year}
  {2021})}\BibitemShut {NoStop}%
\bibitem [{\citenamefont {Jiang}\ \emph {et~al.}(2021)\citenamefont {Jiang},
  \citenamefont {Ding}, \citenamefont {Xi}, \citenamefont {He},\ and\
  \citenamefont {Jiang}}]{24}%
  \BibitemOpen
  \bibfield  {author} {\bibinfo {author} {\bibfnamefont {Z.}~\bibnamefont
  {Jiang}}, \bibinfo {author} {\bibfnamefont {Y.}~\bibnamefont {Ding}},
  \bibinfo {author} {\bibfnamefont {C.}~\bibnamefont {Xi}}, \bibinfo {author}
  {\bibfnamefont {G.}~\bibnamefont {He}},\ and\ \bibinfo {author}
  {\bibfnamefont {C.}~\bibnamefont {Jiang}},\ }\bibfield  {title} {\bibinfo
  {title} {Topological protection of continuous frequency entangled biphoton
  states},\ }\href@noop {} {\bibfield  {journal} {\bibinfo  {journal}
  {Nanophotonics}\ }\textbf {\bibinfo {volume} {10}},\ \bibinfo {pages} {4019}
  (\bibinfo {year} {2021})}\BibitemShut {NoStop}%
\bibitem [{\citenamefont {Jiang}\ \emph {et~al.}(2022)\citenamefont {Jiang},
  \citenamefont {Xi}, \citenamefont {He},\ and\ \citenamefont {Jiang}}]{25}%
  \BibitemOpen
  \bibfield  {author} {\bibinfo {author} {\bibfnamefont {Z.}~\bibnamefont
  {Jiang}}, \bibinfo {author} {\bibfnamefont {C.}~\bibnamefont {Xi}}, \bibinfo
  {author} {\bibfnamefont {G.}~\bibnamefont {He}},\ and\ \bibinfo {author}
  {\bibfnamefont {C.}~\bibnamefont {Jiang}},\ }\bibfield  {title} {\bibinfo
  {title} {Topologically protected energy-time entangled biphoton states in
  photonic crystals},\ }\href@noop {} {\bibfield  {journal} {\bibinfo
  {journal} {Journal of Physics D: Applied Physics}\ }\textbf {\bibinfo
  {volume} {55}},\ \bibinfo {pages} {315104} (\bibinfo {year}
  {2022})}\BibitemShut {NoStop}%
\bibitem [{\citenamefont {Mittal}\ \emph
  {et~al.}(2021{\natexlab{b}})\citenamefont {Mittal}, \citenamefont {Orre},
  \citenamefont {Goldschmidt},\ and\ \citenamefont {Hafezi}}]{26}%
  \BibitemOpen
  \bibfield  {author} {\bibinfo {author} {\bibfnamefont {S.}~\bibnamefont
  {Mittal}}, \bibinfo {author} {\bibfnamefont {V.~V.}\ \bibnamefont {Orre}},
  \bibinfo {author} {\bibfnamefont {E.~A.}\ \bibnamefont {Goldschmidt}},\ and\
  \bibinfo {author} {\bibfnamefont {M.}~\bibnamefont {Hafezi}},\ }\bibfield
  {title} {\bibinfo {title} {Tunable quantum interference using a topological
  source of indistinguishable photon pairs},\ }\href@noop {} {\bibfield
  {journal} {\bibinfo  {journal} {Nature Photonics}\ }\textbf {\bibinfo
  {volume} {15}},\ \bibinfo {pages} {542} (\bibinfo {year}
  {2021}{\natexlab{b}})}\BibitemShut {NoStop}%
\bibitem [{\citenamefont {Gao}\ \emph {et~al.}(2018)\citenamefont {Gao},
  \citenamefont {Xue}, \citenamefont {Yang}, \citenamefont {Lai}, \citenamefont
  {Yu}, \citenamefont {Lin}, \citenamefont {Chong}, \citenamefont {Shvets},\
  and\ \citenamefont {Zhang}}]{27}%
  \BibitemOpen
  \bibfield  {author} {\bibinfo {author} {\bibfnamefont {F.}~\bibnamefont
  {Gao}}, \bibinfo {author} {\bibfnamefont {H.}~\bibnamefont {Xue}}, \bibinfo
  {author} {\bibfnamefont {Z.}~\bibnamefont {Yang}}, \bibinfo {author}
  {\bibfnamefont {K.}~\bibnamefont {Lai}}, \bibinfo {author} {\bibfnamefont
  {Y.}~\bibnamefont {Yu}}, \bibinfo {author} {\bibfnamefont {X.}~\bibnamefont
  {Lin}}, \bibinfo {author} {\bibfnamefont {Y.}~\bibnamefont {Chong}}, \bibinfo
  {author} {\bibfnamefont {G.}~\bibnamefont {Shvets}},\ and\ \bibinfo {author}
  {\bibfnamefont {B.}~\bibnamefont {Zhang}},\ }\bibfield  {title} {\bibinfo
  {title} {Topologically protected refraction of robust kink states in valley
  photonic crystals},\ }\href@noop {} {\bibfield  {journal} {\bibinfo
  {journal} {Nature Physics}\ }\textbf {\bibinfo {volume} {14}},\ \bibinfo
  {pages} {140} (\bibinfo {year} {2018})}\BibitemShut {NoStop}%
\bibitem [{\citenamefont {Yang}\ \emph {et~al.}(2020)\citenamefont {Yang},
  \citenamefont {Yamagami}, \citenamefont {Yu}, \citenamefont {Pitchappa},
  \citenamefont {Webber}, \citenamefont {Zhang}, \citenamefont {Fujita},
  \citenamefont {Nagatsuma},\ and\ \citenamefont {Singh}}]{28}%
  \BibitemOpen
  \bibfield  {author} {\bibinfo {author} {\bibfnamefont {Y.}~\bibnamefont
  {Yang}}, \bibinfo {author} {\bibfnamefont {Y.}~\bibnamefont {Yamagami}},
  \bibinfo {author} {\bibfnamefont {X.}~\bibnamefont {Yu}}, \bibinfo {author}
  {\bibfnamefont {P.}~\bibnamefont {Pitchappa}}, \bibinfo {author}
  {\bibfnamefont {J.}~\bibnamefont {Webber}}, \bibinfo {author} {\bibfnamefont
  {B.}~\bibnamefont {Zhang}}, \bibinfo {author} {\bibfnamefont
  {M.}~\bibnamefont {Fujita}}, \bibinfo {author} {\bibfnamefont
  {T.}~\bibnamefont {Nagatsuma}},\ and\ \bibinfo {author} {\bibfnamefont
  {R.}~\bibnamefont {Singh}},\ }\bibfield  {title} {\bibinfo {title} {Terahertz
  topological photonics for on-chip communication},\ }\href@noop {} {\bibfield
  {journal} {\bibinfo  {journal} {Nature Photonics}\ }\textbf {\bibinfo
  {volume} {14}},\ \bibinfo {pages} {446} (\bibinfo {year} {2020})}\BibitemShut
  {NoStop}%
\bibitem [{\citenamefont {Shalaev}\ \emph {et~al.}(2019)\citenamefont
  {Shalaev}, \citenamefont {Walasik}, \citenamefont {Tsukernik}, \citenamefont
  {Xu},\ and\ \citenamefont {Litchinitser}}]{29}%
  \BibitemOpen
  \bibfield  {author} {\bibinfo {author} {\bibfnamefont {M.~I.}\ \bibnamefont
  {Shalaev}}, \bibinfo {author} {\bibfnamefont {W.}~\bibnamefont {Walasik}},
  \bibinfo {author} {\bibfnamefont {A.}~\bibnamefont {Tsukernik}}, \bibinfo
  {author} {\bibfnamefont {Y.}~\bibnamefont {Xu}},\ and\ \bibinfo {author}
  {\bibfnamefont {N.~M.}\ \bibnamefont {Litchinitser}},\ }\bibfield  {title}
  {\bibinfo {title} {Robust topologically protected transport in photonic
  crystals at telecommunication wavelengths},\ }\href@noop {} {\bibfield
  {journal} {\bibinfo  {journal} {Nature Nanotechnology}\ }\textbf {\bibinfo
  {volume} {14}},\ \bibinfo {pages} {31} (\bibinfo {year} {2019})}\BibitemShut
  {NoStop}%
\bibitem [{\citenamefont {Wang}\ \emph
  {et~al.}(2022{\natexlab{a}})\citenamefont {Wang}, \citenamefont {Sun},
  \citenamefont {He}, \citenamefont {Tang}, \citenamefont {An}, \citenamefont
  {Wang}, \citenamefont {Du}, \citenamefont {Zhang}, \citenamefont {Yuan},
  \citenamefont {He} \emph {et~al.}}]{30}%
  \BibitemOpen
  \bibfield  {author} {\bibinfo {author} {\bibfnamefont {H.}~\bibnamefont
  {Wang}}, \bibinfo {author} {\bibfnamefont {L.}~\bibnamefont {Sun}}, \bibinfo
  {author} {\bibfnamefont {Y.}~\bibnamefont {He}}, \bibinfo {author}
  {\bibfnamefont {G.}~\bibnamefont {Tang}}, \bibinfo {author} {\bibfnamefont
  {S.}~\bibnamefont {An}}, \bibinfo {author} {\bibfnamefont {Z.}~\bibnamefont
  {Wang}}, \bibinfo {author} {\bibfnamefont {Y.}~\bibnamefont {Du}}, \bibinfo
  {author} {\bibfnamefont {Y.}~\bibnamefont {Zhang}}, \bibinfo {author}
  {\bibfnamefont {L.}~\bibnamefont {Yuan}}, \bibinfo {author} {\bibfnamefont
  {X.}~\bibnamefont {He}}, \emph {et~al.},\ }\bibfield  {title} {\bibinfo
  {title} {Asymmetric topological valley edge states on silicon-on-insulator
  platform},\ }\href@noop {} {\bibfield  {journal} {\bibinfo  {journal} {Laser
  \& Photonics Reviews}\ }\textbf {\bibinfo {volume} {16}},\ \bibinfo {pages}
  {2100631} (\bibinfo {year} {2022}{\natexlab{a}})}\BibitemShut {NoStop}%
\bibitem [{\citenamefont {He}\ \emph {et~al.}(2019{\natexlab{b}})\citenamefont
  {He}, \citenamefont {Liang}, \citenamefont {Yuan}, \citenamefont {Qiu},
  \citenamefont {Chen}, \citenamefont {Zhao},\ and\ \citenamefont {Dong}}]{31}%
  \BibitemOpen
  \bibfield  {author} {\bibinfo {author} {\bibfnamefont {X.-T.}\ \bibnamefont
  {He}}, \bibinfo {author} {\bibfnamefont {E.-T.}\ \bibnamefont {Liang}},
  \bibinfo {author} {\bibfnamefont {J.-J.}\ \bibnamefont {Yuan}}, \bibinfo
  {author} {\bibfnamefont {H.-Y.}\ \bibnamefont {Qiu}}, \bibinfo {author}
  {\bibfnamefont {X.-D.}\ \bibnamefont {Chen}}, \bibinfo {author}
  {\bibfnamefont {F.-L.}\ \bibnamefont {Zhao}},\ and\ \bibinfo {author}
  {\bibfnamefont {J.-W.}\ \bibnamefont {Dong}},\ }\bibfield  {title} {\bibinfo
  {title} {A silicon-on-insulator slab for topological valley transport},\
  }\href@noop {} {\bibfield  {journal} {\bibinfo  {journal} {Nature
  Communications}\ }\textbf {\bibinfo {volume} {10}},\ \bibinfo {pages} {872}
  (\bibinfo {year} {2019}{\natexlab{b}})}\BibitemShut {NoStop}%
\bibitem [{\citenamefont {Zeng}\ \emph {et~al.}(2020)\citenamefont {Zeng},
  \citenamefont {Chattopadhyay}, \citenamefont {Zhu}, \citenamefont {Qiang},
  \citenamefont {Li}, \citenamefont {Jin}, \citenamefont {Li}, \citenamefont
  {Davies}, \citenamefont {Linfield}, \citenamefont {Zhang} \emph
  {et~al.}}]{32}%
  \BibitemOpen
  \bibfield  {author} {\bibinfo {author} {\bibfnamefont {Y.}~\bibnamefont
  {Zeng}}, \bibinfo {author} {\bibfnamefont {U.}~\bibnamefont {Chattopadhyay}},
  \bibinfo {author} {\bibfnamefont {B.}~\bibnamefont {Zhu}}, \bibinfo {author}
  {\bibfnamefont {B.}~\bibnamefont {Qiang}}, \bibinfo {author} {\bibfnamefont
  {J.}~\bibnamefont {Li}}, \bibinfo {author} {\bibfnamefont {Y.}~\bibnamefont
  {Jin}}, \bibinfo {author} {\bibfnamefont {L.}~\bibnamefont {Li}}, \bibinfo
  {author} {\bibfnamefont {A.~G.}\ \bibnamefont {Davies}}, \bibinfo {author}
  {\bibfnamefont {E.~H.}\ \bibnamefont {Linfield}}, \bibinfo {author}
  {\bibfnamefont {B.}~\bibnamefont {Zhang}}, \emph {et~al.},\ }\bibfield
  {title} {\bibinfo {title} {Electrically pumped topological laser with valley
  edge modes},\ }\href@noop {} {\bibfield  {journal} {\bibinfo  {journal}
  {Nature}\ }\textbf {\bibinfo {volume} {578}},\ \bibinfo {pages} {246}
  (\bibinfo {year} {2020})}\BibitemShut {NoStop}%
\bibitem [{\citenamefont {Zhong}\ \emph {et~al.}(2020)\citenamefont {Zhong},
  \citenamefont {Li}, \citenamefont {Song}, \citenamefont {Kartashov},
  \citenamefont {Zhang}, \citenamefont {Zhang},\ and\ \citenamefont
  {Chen}}]{33}%
  \BibitemOpen
  \bibfield  {author} {\bibinfo {author} {\bibfnamefont {H.}~\bibnamefont
  {Zhong}}, \bibinfo {author} {\bibfnamefont {Y.}~\bibnamefont {Li}}, \bibinfo
  {author} {\bibfnamefont {D.}~\bibnamefont {Song}}, \bibinfo {author}
  {\bibfnamefont {Y.~V.}\ \bibnamefont {Kartashov}}, \bibinfo {author}
  {\bibfnamefont {Y.}~\bibnamefont {Zhang}}, \bibinfo {author} {\bibfnamefont
  {Y.}~\bibnamefont {Zhang}},\ and\ \bibinfo {author} {\bibfnamefont
  {Z.}~\bibnamefont {Chen}},\ }\bibfield  {title} {\bibinfo {title}
  {Topological valley hall edge state lasing},\ }\href@noop {} {\bibfield
  {journal} {\bibinfo  {journal} {Laser \& Photonics Reviews}\ }\textbf
  {\bibinfo {volume} {14}},\ \bibinfo {pages} {2000001} (\bibinfo {year}
  {2020})}\BibitemShut {NoStop}%
\bibitem [{\citenamefont {Ma}\ \emph {et~al.}(2019)\citenamefont {Ma},
  \citenamefont {Xi},\ and\ \citenamefont {Sun}}]{34}%
  \BibitemOpen
  \bibfield  {author} {\bibinfo {author} {\bibfnamefont {J.}~\bibnamefont
  {Ma}}, \bibinfo {author} {\bibfnamefont {X.}~\bibnamefont {Xi}},\ and\
  \bibinfo {author} {\bibfnamefont {X.}~\bibnamefont {Sun}},\ }\bibfield
  {title} {\bibinfo {title} {Topological photonic integrated circuits based on
  valley kink states},\ }\href@noop {} {\bibfield  {journal} {\bibinfo
  {journal} {Laser \& Photonics Reviews}\ }\textbf {\bibinfo {volume} {13}},\
  \bibinfo {pages} {1900087} (\bibinfo {year} {2019})}\BibitemShut {NoStop}%
\bibitem [{\citenamefont {Gu}\ \emph {et~al.}(2021)\citenamefont {Gu},
  \citenamefont {Yuan}, \citenamefont {Zhao}, \citenamefont {Ji}, \citenamefont
  {Liu}, \citenamefont {Fang}, \citenamefont {Gan},\ and\ \citenamefont
  {Zhao}}]{35}%
  \BibitemOpen
  \bibfield  {author} {\bibinfo {author} {\bibfnamefont {L.}~\bibnamefont
  {Gu}}, \bibinfo {author} {\bibfnamefont {Q.}~\bibnamefont {Yuan}}, \bibinfo
  {author} {\bibfnamefont {Q.}~\bibnamefont {Zhao}}, \bibinfo {author}
  {\bibfnamefont {Y.}~\bibnamefont {Ji}}, \bibinfo {author} {\bibfnamefont
  {Z.}~\bibnamefont {Liu}}, \bibinfo {author} {\bibfnamefont {L.}~\bibnamefont
  {Fang}}, \bibinfo {author} {\bibfnamefont {X.}~\bibnamefont {Gan}},\ and\
  \bibinfo {author} {\bibfnamefont {J.}~\bibnamefont {Zhao}},\ }\bibfield
  {title} {\bibinfo {title} {A topological photonic ring-resonator for on-chip
  channel filters},\ }\href@noop {} {\bibfield  {journal} {\bibinfo  {journal}
  {Journal of Lightwave Technology}\ }\textbf {\bibinfo {volume} {39}},\
  \bibinfo {pages} {5069} (\bibinfo {year} {2021})}\BibitemShut {NoStop}%
\bibitem [{\citenamefont {Barik}\ \emph {et~al.}(2020)\citenamefont {Barik},
  \citenamefont {Karasahin}, \citenamefont {Mittal}, \citenamefont {Waks},\
  and\ \citenamefont {Hafezi}}]{36}%
  \BibitemOpen
  \bibfield  {author} {\bibinfo {author} {\bibfnamefont {S.}~\bibnamefont
  {Barik}}, \bibinfo {author} {\bibfnamefont {A.}~\bibnamefont {Karasahin}},
  \bibinfo {author} {\bibfnamefont {S.}~\bibnamefont {Mittal}}, \bibinfo
  {author} {\bibfnamefont {E.}~\bibnamefont {Waks}},\ and\ \bibinfo {author}
  {\bibfnamefont {M.}~\bibnamefont {Hafezi}},\ }\bibfield  {title} {\bibinfo
  {title} {Chiral quantum optics using a topological resonator},\ }\href@noop
  {} {\bibfield  {journal} {\bibinfo  {journal} {Physical Review B}\ }\textbf
  {\bibinfo {volume} {101}},\ \bibinfo {pages} {205303} (\bibinfo {year}
  {2020})}\BibitemShut {NoStop}%
\bibitem [{\citenamefont {Lugiato}\ and\ \citenamefont {Lefever}(1987)}]{37}%
  \BibitemOpen
  \bibfield  {author} {\bibinfo {author} {\bibfnamefont {L.~A.}\ \bibnamefont
  {Lugiato}}\ and\ \bibinfo {author} {\bibfnamefont {R.}~\bibnamefont
  {Lefever}},\ }\bibfield  {title} {\bibinfo {title} {Spatial dissipative
  structures in passive optical systems},\ }\href@noop {} {\bibfield  {journal}
  {\bibinfo  {journal} {Physical Review Letters}\ }\textbf {\bibinfo {volume}
  {58}},\ \bibinfo {pages} {2209} (\bibinfo {year} {1987})}\BibitemShut
  {NoStop}%
\bibitem [{\citenamefont {Chembo}\ and\ \citenamefont {Menyuk}(2013)}]{38}%
  \BibitemOpen
  \bibfield  {author} {\bibinfo {author} {\bibfnamefont {Y.~K.}\ \bibnamefont
  {Chembo}}\ and\ \bibinfo {author} {\bibfnamefont {C.~R.}\ \bibnamefont
  {Menyuk}},\ }\bibfield  {title} {\bibinfo {title} {Spatiotemporal
  lugiato-lefever formalism for kerr-comb generation in whispering-gallery-mode
  resonators},\ }\href@noop {} {\bibfield  {journal} {\bibinfo  {journal}
  {Physical Review A}\ }\textbf {\bibinfo {volume} {87}},\ \bibinfo {pages}
  {053852} (\bibinfo {year} {2013})}\BibitemShut {NoStop}%
\bibitem [{\citenamefont {Gaeta}\ \emph {et~al.}(2019)\citenamefont {Gaeta},
  \citenamefont {Lipson},\ and\ \citenamefont {Kippenberg}}]{39}%
  \BibitemOpen
  \bibfield  {author} {\bibinfo {author} {\bibfnamefont {A.~L.}\ \bibnamefont
  {Gaeta}}, \bibinfo {author} {\bibfnamefont {M.}~\bibnamefont {Lipson}},\ and\
  \bibinfo {author} {\bibfnamefont {T.~J.}\ \bibnamefont {Kippenberg}},\
  }\bibfield  {title} {\bibinfo {title} {Photonic-chip-based frequency combs},\
  }\href@noop {} {\bibfield  {journal} {\bibinfo  {journal} {Nature Photonics}\
  }\textbf {\bibinfo {volume} {13}},\ \bibinfo {pages} {158} (\bibinfo {year}
  {2019})}\BibitemShut {NoStop}%
\bibitem [{\citenamefont {Wang}\ \emph
  {et~al.}(2022{\natexlab{b}})\citenamefont {Wang}, \citenamefont {Tang},
  \citenamefont {He}, \citenamefont {Wang}, \citenamefont {Li}, \citenamefont
  {Sun}, \citenamefont {Zhang}, \citenamefont {Yuan}, \citenamefont {Dong},\
  and\ \citenamefont {Su}}]{40}%
  \BibitemOpen
  \bibfield  {author} {\bibinfo {author} {\bibfnamefont {H.}~\bibnamefont
  {Wang}}, \bibinfo {author} {\bibfnamefont {G.}~\bibnamefont {Tang}}, \bibinfo
  {author} {\bibfnamefont {Y.}~\bibnamefont {He}}, \bibinfo {author}
  {\bibfnamefont {Z.}~\bibnamefont {Wang}}, \bibinfo {author} {\bibfnamefont
  {X.}~\bibnamefont {Li}}, \bibinfo {author} {\bibfnamefont {L.}~\bibnamefont
  {Sun}}, \bibinfo {author} {\bibfnamefont {Y.}~\bibnamefont {Zhang}}, \bibinfo
  {author} {\bibfnamefont {L.}~\bibnamefont {Yuan}}, \bibinfo {author}
  {\bibfnamefont {J.}~\bibnamefont {Dong}},\ and\ \bibinfo {author}
  {\bibfnamefont {Y.}~\bibnamefont {Su}},\ }\bibfield  {title} {\bibinfo
  {title} {Ultracompact topological photonic switch based on
  valley-vortex-enhanced high-efficiency phase shift},\ }\href@noop {}
  {\bibfield  {journal} {\bibinfo  {journal} {Light: Science \& Applications}\
  }\textbf {\bibinfo {volume} {11}},\ \bibinfo {pages} {292} (\bibinfo {year}
  {2022}{\natexlab{b}})}\BibitemShut {NoStop}%
\bibitem [{\citenamefont {Wang}\ \emph {et~al.}(2020)\citenamefont {Wang},
  \citenamefont {Wang},\ and\ \citenamefont {Zhang}}]{41}%
  \BibitemOpen
  \bibfield  {author} {\bibinfo {author} {\bibfnamefont {W.}~\bibnamefont
  {Wang}}, \bibinfo {author} {\bibfnamefont {L.}~\bibnamefont {Wang}},\ and\
  \bibinfo {author} {\bibfnamefont {W.}~\bibnamefont {Zhang}},\ }\bibfield
  {title} {\bibinfo {title} {Advances in soliton microcomb generation},\
  }\href@noop {} {\bibfield  {journal} {\bibinfo  {journal} {Advanced
  Photonics}\ }\textbf {\bibinfo {volume} {2}},\ \bibinfo {pages} {034001}
  (\bibinfo {year} {2020})}\BibitemShut {NoStop}%
\end{thebibliography}


%

\end{document}


\renewcommand{\thefigure}{S\arabic{figure}}
\renewcommand{\theequation}{S\arabic{equation}}
\renewcommand{\thetable}{S\arabic{table}}

\preprint{APS}

\title{Supplementary Materials: Topological dissipative Kerr soliton combs in a valley photonic crystal resonator}

\author{Zhen Jiang$^{1,2}$} \author{Lefeng Zhou$^{1}$}%
\author{Wei Li$^{3}$}%
\author{Yudong Li$^{3}$}%
\author{Liangsen Feng$^{3}$}%
\author{Tengfei Wu$^{3}$}%
\author{Chun Jiang$^{1}$}  \email{cjiang@sjtu.edu.cn}	
\author{Guangqiang He$^{1,2}$}  \email{gqhe@sjtu.edu.cn}	
\affiliation{%
 $^1$State Key Laboratory of Advanced Optical Communication Systems and Networks, Department of Electronic Engineering, Shanghai Jiao Tong University, Shanghai 200240, China\\
 $^2$SJTU Pinghu Institute of Intelligent Optoelectronics, Department of Electronic Engineering, Shanghai Jiao Tong University, Shanghai 200240, China \\
 $^3$Science and Technology on Metrology and Calibration Laboratory, Changcheng Institute of Metrology $\rm \&$ Measurement, Aviation Industry Corporation of China, Beijing 100095, China
}%
	
\maketitle

\section{Numerical simulation}
We use the commercial software COMSOL Multiphysics to simulate the FWM process in the  $\rm Si_{3}N_{4}$ topological resonator (Fig. 3 in the main text). Here we consider the field profiles of transverse electric (TE) polarization modes. We choose one resonator mode (indexed by the number $\mu=0$) at the frequency of $v_p=370.6$7 THz as the pump mode, where the neighboring two resonator modes ($\mu=1,-1$) are selected as the signal ($v_s=371.13 $THz) and idler ($v_i=370.21$ THz) frequencies, respectively. The third-order nonlinearity of  $\rm Si_{3}N_{4}$ $\chi^{(3)}=2.5\times10^{-19}\mathrm{m}^2/\mathrm{V}^2$ is used to introduce the nonlinear elements. The FWM process conducted by third-order nonlinear polarizations of $\rm Si_{3}N_{4}$ can be described by
\begin{equation}
\begin{aligned}
\mathbf{P}_{p}\left(\omega_{s}+\omega_{i}-\omega_{p}\right)=6\varepsilon_{0}\chi^{(3)}E_{s}E_{i}E_{p}^{*}, \\ 
\mathbf{P}_{s}\left(\omega_{p}+\omega_{p}-\omega_{i}\right)=3\varepsilon_{0}\chi^{(3)}E_{p}E_{p}E_{i}^{*}, \\ 
\mathbf{P}_{i}\left(\omega_{p}+\omega_{p}-\omega_{s}\right)=3\varepsilon_{0}\chi^{(3)}E_{p}E_{p}E_{s}^{*}, \\
\label{eq:S1}
\end{aligned}
\end{equation}	
where $\mathbf{P}_{p,s,i}$ and $E_{p,s,i}$  are the polarizations and electric field of the pump, signal, and idler. In the numerical model, we use sources with $E=E_x+iE_y$ set at the input port of the bus waveguide to emulate right circularly polarized light. And the input power of the pump and signal are set as $|E_p|=10^4$ $\rm {V/m}$ and.
$|E_s|=10^3$ $\rm {V/m}$  respectively, with no input for the idler. As shown in Fig. 3 in the main text, the excitation of idler modes indicates that nonlinear coupling between the pump and signal model at the idler frequency is driven by the nonlinear polarization of electromagnetic models at the pump and signal frequencies. 

\section{Valley topological photonics (VPCs)}
\begin{figure*}
\centering
\includegraphics[width=0.9\textwidth]{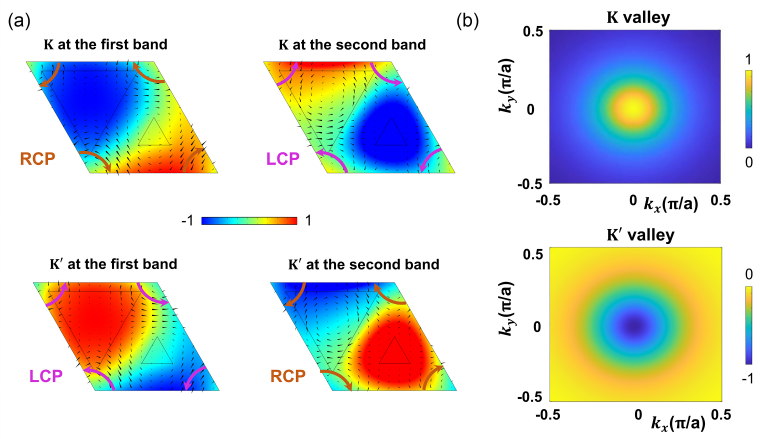}
\caption{(a) Simulated magnetic field profiles $H_z$ at the $\rm K$ and $\rm K'$  valleys for the first and second bands of the $\rm VPC_2$. (b) Normalized simulated Berry curvatures near the  $\rm K$ and $\rm K'$ valleys at the first band.}
 \label{fig:S1}
\end{figure*}

For unperturbed unit cells with $C_6$ lattice symmetry, there exist degenerate Dirac points at the $\rm K$ and $\rm K'$ valleys. The effective Hamiltonian in the vicinity of $\rm K$ ($\rm K'$) point is described as~\cite{s1,s2,s3}

\begin{equation}
H_{K/K'}^{(0)}=v_{D}\bigl(\sigma_{x}\delta k_{x}+\sigma_{y}\delta k_{y}\bigr),
\label{eq:S2}
\end{equation}	
where $v_{D}$ is the group velocity, $\sigma_{x}$ and $\sigma_{y}$ are the Pauli matrices, $\delta\vec{k}=\vec{k}-\vec{k}_{K/K'}$ denotes the deviation of the wavevector. With the distortion of the unit cell ($d_1 \neq d_2$), the Hamiltonian can be rewritten as 
\begin{equation}
H_{K/K'}=v_D\bigl(\sigma_x\delta k_x+\sigma_y\delta k_y\bigr)+v_Dm\sigma_z,
\label{eq:S3}
\end{equation}	
where m denotes the strength of the symmetry-breaking perturbation. The simulated magnetic field profiles $H_z$ at the $\rm K$ and $\rm K'$ valleys for the first (367.01 THz) and second (409.77 THz) bands of the $\rm VPC_2$ are shown in Fig.~\ref{fig:S1}(a). It reveals that the modes at the $\rm K$ and $\rm K'$ valleys show opposite circular polarizations, to be specific, left-handed circular polarization (LCP) and right-handed circular polarization (RCP) respectively. The normalized simulated Berry curvatures are plotted in Fig.~\ref{fig:S1}(b). To characterize the topological properties of VPCs, we calculate the valley Chern numbers of VPCs by~\cite{s1,s4}
\begin{equation}
C_{K/K^{\prime}}=\frac{1}{2\pi}\int_{HBZ}\Omega_{K/K^{\prime}}(\delta\vec{k})dS=\pm1/2,
\label{eq:S4}
\end{equation}	
where $\Omega_{K/K'}$ is the Berry curvature, and this integration region contains half of the Brillouin zone. Therefore, the difference of valley Chern number of the system is calculated as $|C_{K/K'}|=1$, leading to the topological nature of VPCs.
\section{Topological edge states}
\begin{figure*}[ht]
\centering
\includegraphics[width=0.8\textwidth]{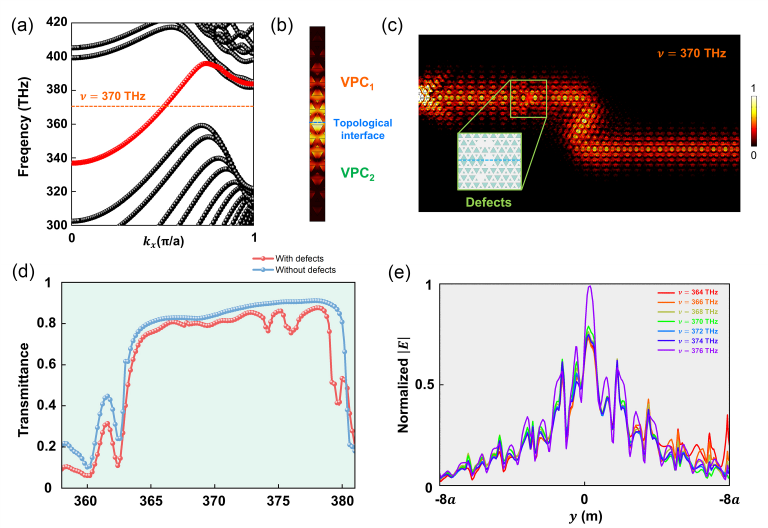}
\caption{(a) Calculated dispersion curves of VPCs. (b) Field distribution for the valley kink state around the interface with $k_x=0.7 (\pi/a$), respectively. (c) Simulated field profiles of valley kink states along the “Z” shaped topological waveguide with defects at 370 THz. (d) Linear transmittances of kink states. (e) Normalized electric field distributions $\rm |E|$ along y axis.}
    \label{fig:S2}
\end{figure*}

Topologically protected edge states can be observed at the interface between $\rm VPC_1$ and $\rm VPC_2$, these edge states are also called valley kink states~\cite{s5}. Fig.~\ref{fig:S2}(a) shows the calculated dispersion curves of the valley kink state with $k_x>0$, where the red curve denotes the valley kink state. The field distribution for the valley kink states around the interface with $k_x=0.7 (\pi/a)$ is plotted in Fig.~\ref{fig:S2}(b). To verify the robustness of such edge states, we design a “Z” shaped topological waveguide with several random lacking of holes around the interface. The simulated field profile of valley kink states at the frequency of 370 THz along the interface is illustrated in Fig.~\ref{fig:S2}(c). It shows that the light can smoothly pass through the defects and sharp corners, leading to immunity to these defects. We further discuss the linear transmittances of the valley edge states at the “Z” shaped interfaces with or without defects. Two detector dipoles are set to monitor the power of the input and output ports. As shown in Fig.~\ref{fig:S2}(d), an abrupt decrease in the transmittance has not been observed. This behavior also gives clear evidence of robustness against defects. To discuss the localization of valley kink states, we plot the normalized electric field distributions $\rm |E|$ along y axis with a series of frequencies. As depicted in Fig.~\ref{fig:S2}(e), the electric field is concentrated at the interface between two VPCs, which implies the good light confinement of valley kink states. 

\section{Dispersion engineering}
To describe the light propagation of topological edge states, the dispersion characteristics of these states are studied. The propagation constant $\beta$ can also be expanded around frequency $\omega_0$ in a Taylor series:
\begin{equation}
\beta=\beta_0+\beta_1(\omega-\omega_0)+\beta_2\frac{(\omega-\omega_0)^2}{2}+\cdots,
\label{eq:S5}
\end{equation}	
where the expansion term is $\beta_{i}=d^{i}k/d\omega^{i}$ at $\omega=\omega_0$. The first-order term $\beta_1$ is linked to the group velocity of the topological edge state: $\beta_{1}=1/\nu_{g}$, where the group velocity is given by $v_{g}=d\omega/dk$. The second term, $\beta_2$ so-called group velocity dispersion (GVD) can also be used to identify the dispersion characteristics of topological edge states, with normal dispersion ($\beta_2>0$) and anomalous dispersion ($\beta_2<0$). 
\begin{figure*}[ht]
\centering
\includegraphics[width=0.8\textwidth]{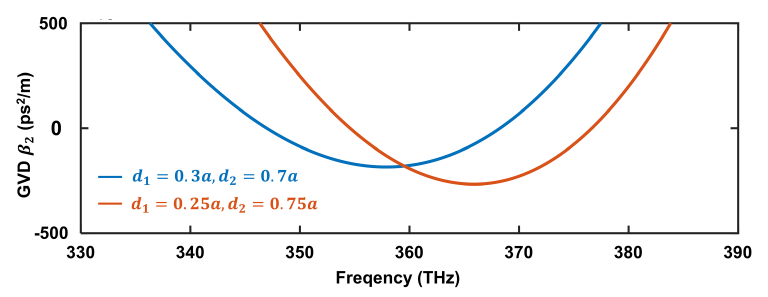}
\caption{Simulated GVD curves for topological edge states with different structure parameters: $d_1=0.3a$, $d_2=0.7a$ and $d_1=0.25a$, $d_2=0.75a$, respectively.}
    \label{fig:S3}
\end{figure*}
Here we calculate the GVD curve of the topological edge state for the wavevector $k_x<0$, using different structure parameters: $d_1=0.3a$, $d_2=0.7a$ and $d_1=0.25a$, $d_2=0.75a$, respectively (Fig.~\ref{fig:S3}). It exhibits anomalous dispersion characteristics around the pump frequency of 370 THz for the structure parameters with $d_1=0.25a$, $d_2=0.75a$. It is worth mentioning that the $\beta_i$ is highly related to the integrated dispersion with $\beta_{1}=2\pi/LD_{1}$ and $\beta_{2}=-2\pi D_{2}/LD_{1}^{3}$, respectively. We can calculate the dispersion $D_2=9.60$ GHz, which is approximately equal to the dispersion of the topological resonator extracted from the simulated transmission. Therefore, the topological resonator can be approximatively considered as a single-ring resonator in the dynamic evolution simulation of Kerr soliton combs.

\section{Coupling between waveguides and topological resonators}
\begin{figure*}[ht]
\centering
\includegraphics[width=0.9\textwidth]{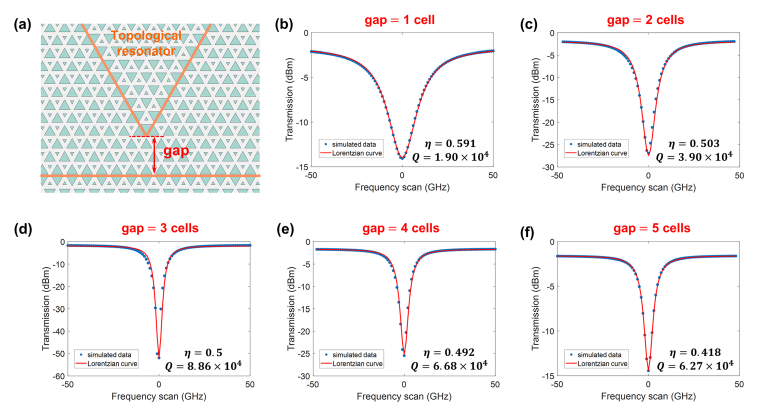}
\caption{(a) Close-up image of coupling region between the bus waveguide and the topological resonator. Lorentzian fittings of the simulated resonant dips at the pump frequency $v_p$ around 370.6 THz with the gaps of (b) 1 cell, (c) 2 cells, (d) 3 cells, (e) 4 cells, and (f) 5 cells, respectively.}
    \label{fig:S4}
\end{figure*}

Here we use a straight topological bus waveguide to couple the pump mode into the topological resonator, and then, guide the generated combs from the cavity to the bus waveguide. To assess the couplings between a straight waveguide and a triangular resonator, we design several configurations with a series of gap widths. As shown in Fig.~\ref{fig:S4}(a), the values of the gap width are set from 1 cell to 5 cells. Note that the total Q-factor is given by $1⁄Q=1⁄Q_{in} +1⁄Q_{ex}$, where $Q_{in}$ and $Q_{ex}$ are the intrinsic quality factor and external quality factor respectively. The $Q_{in}$ and $Q_{ex}$ can be calculated as $Q_{in}=\omega/\alpha v_{g}=\omega/\kappa_{in}\approx\omega/(1-e^{-\alpha L})FSR$ and $Q_{ex}=\omega/\kappa_{ex}\approx\omega/(1-|t_{1}|^{2})FSR$, where $\alpha$, $v_g$, and $L$ represent the roundtrip loss, the group velocity, and the roundtrip length of the topological cavity, respectively. The parameters $\kappa_{in}$, $\kappa_{ex}$, and $\kappa$ denote the intrinsic loss, the external loss, and the total energy loss rate of the resonator. 

The Lorentzian fittings of the simulated resonant dips at the pump frequency $v_p$ around 370.6 THz are plotted in Fig.~\ref{fig:S4}(b)-(f). The result shows that the topological resonator with the gap width of 3 cells (Fig.~\ref{fig:S4}(d)) satisfies $\eta=1⁄2$, leading to the critical coupling. For the critical coupling, the external loss is equal to the intrinsic loss with $\kappa_{in}=\kappa_{ex}=1.35\times10^7$ rad/s. Therefore, the intrinsic quality factor and external quality factor are $Q_{in}=Q_{ex}=1.772\times10^5$, leading to a value of a total Q-factor with $Q=8.86\times10^4$. We can also find that the topological resonators are over-coupling ($\eta>1⁄2$) when the gap widths are small than 3 cells. Correspondingly, the topological resonators are under coupling ($\eta<1⁄2$) when the gap widths are large than 3 cells.
\section{Topological resonator modes}
\begin{figure*}[ht]
\centering
\includegraphics[width=0.8\textwidth]{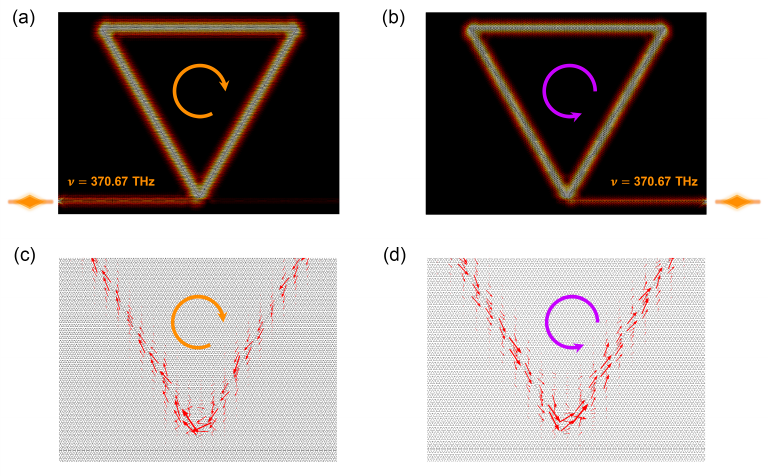}
\caption{(a) Simulated field profiles and (c) corresponding Poynting vector distributions of resonator modes at the pump frequency with the excitation at the left of the waveguide. (b) Simulated field profiles and (d) corresponding Poynting vector distributions for the excitation at the right of the waveguide.}
    \label{fig:S5}
\end{figure*}

Considering the light propagation behavior of valley kink states in the topological resonator, we simulate the field profiles and energy flow of resonator modes at the pump frequency. We pump the bus waveguide in the opposite direction. As shown in Fig.~\ref{fig:S5}(a), the pump mode excited at the left of the waveguide is coupled into the topological resonator. Its corresponding Poynting vector distributions plotted in Fig.~\ref{fig:S5}(c) reveal that the energy travels clockwise inside the triangular cavity. And then, the energy is coupled out from the cavity to the right of the waveguide. Due to time-reversal symmetry, the pump mode excited in the opposite direction excites an anticlockwise rotation mode inside the cavity (Fig.~\ref{fig:S5}(b), (d)).

\section{Numerical simulation of DKS combs in topological resonators}
Here we theoretically analyze the nonlinear evolution of the pump field in the topological resonator. The Lugiato–Lefever equation (LLE)~\cite{s6} can be concluded from the nonlinear Schrödinger equation (NSE) with boundary conditions of the resonators, where the NSE has the form:
\begin{equation}
\frac{\partial}{\partial z}A^{(m)}+\frac{\alpha}{2}A^{(m)}+i\frac{\beta_{2}}{2}\frac{\partial^{2}}{\partial r^{2}}A^{(m)}=i\gamma\big|A^{(m)}\big|^{2}A^{(m)},
\label{eq:S6}
\end{equation}	
where m denotes the $m$-th roundtrips travel of the light field, $A=A(z,t)$ is the field envelope with the propagation distance $z$ in resonators, $T$ is the fast time variable describing the waveform. $\alpha$, $\beta_{2}$, and $\gamma$ are the roundtrip loss, the dispersion term, and the nonlinear coefficient, respectively. The boundary condition can be written as
\begin{equation}
A^{(m+1)}(0,T)=\sqrt{\Theta}A_i+\sqrt{1-\Theta}\exp(-i\delta_0)A^{(m)}(L,T),
\label{eq:S7}
\end{equation}	
in which $A_i$ is the light field of the pump, $\Theta$ denotes the power coupling coefficient, $\delta_0$ is the detuning of the resonance frequency, and $L$ is length of the topological resonator. $\frac{\partial}{\partial z}A^{(m)}$ takes an approximate form of 
\begin{equation}
\frac{\partial}{\partial z}A^{(m)}(z,T)\bigg|_{z=0}=\frac{A^{(m)}(L,T)-A^{(m)}(0,T)}{L}.
\label{eq:S8}
\end{equation}	

When the light field does not pass through the coupling region between the topological resonator and the straight waveguide, the nonlinear evolution is governed by NSE. Substitute the Eq.~\ref{eq:S8} into Eq.~\ref{eq:S6}, we can get
\begin{equation}
A^{(m)}(L,T)=A^{(m)}(0,T)+L\left(-\frac{\alpha}{2}A^{(m)}(0,T)-\frac{i\beta_{2}}{2}\frac{\partial^{2}}{\partial T^{2}}A^{(m)}(0,T)+i\gamma\Big|A^{(m)}(0,T)\Big|^{2}A^{(m)}(0,T)\Big).\right. 
\label{eq:S9}
\end{equation}	
The NSE reveals how the light field evolves after the propagation length $L$.
Next, we take the coupling region into consideration. In this step, we take two assumptions: first, the power coupling coefficient is small, that is, $\Theta\ll 1$; second, the detuning compared with the FSR is very small, namely $\delta_{0}\ll2\pi$. Therefore, the Eq.~\ref{eq:S7} can be reduced to 
\begin{equation}
A^{(m+1)}(0,T)=\sqrt{\Theta}A_i+\left(1-\frac{\Theta}{2}-i\delta_0\right)A^{(m)}(L,T).
\label{eq:S10}
\end{equation}	

Since the term m is a discrete value, it leads to complexity for derivations. A parameter $t_R$, the so-called slow time variable, to replace the term m. When m increases to $m+1$, it is the equivalent of increasing time $t_R$ (roundtrip time) for the current time $\tau$. We can obtain the relation
\begin{equation}
\frac{\partial}{\partial\tau}A(\tau,T)=\frac{A^{(m+1)}(0,T)-A^{(m)}(0,T)}{t_{R}}.
\label{eq:S11}
\end{equation}	

Substitute the Eq.~\ref{eq:S9} and Eq.~\ref{eq:S10} into Eq.~\ref{eq:S11}, the term can be rewritten as 
\begin{equation}
t_R\frac{\partial}{\partial\tau}A=-\left(\frac{\alpha L+\Theta}{2}+i\delta_0\right)A-iL\frac{\beta_2}{2}\frac{\partial^2}{\partial T^2}A+iL\gamma|A|^2A+\sqrt{\Theta}A_i.
\label{eq:S12}
\end{equation}	
Eq.~\ref{eq:S12} is the first form of LLE. To obtain the second form of the equation, new parameters are introduced to endow the physical meaning of the model. First, the roundtrip time $t_R$ is related to the FSR by $t_R=1⁄FSR$. The normalized losses are defined as
\begin{equation}
\kappa_{in}=\alpha L\cdot FSR,\quad\kappa_{ex}=\Theta\cdot FSR,\quad\kappa=\kappa_{in}+\kappa_{ex},
\label{eq:S13}
\end{equation}	
where $\kappa_{in}$, $\kappa_{ex}$, and $\kappa$ denote the intrinsic loss, the external loss, and the total energy loss rate of the resonator. The parameter $\kappa$ is also related to the linewidth of resonator modes. Then, we can define the normalized detuning as
\begin{equation}
\delta_{0}=\beta_{1}L\big(\omega_{0}-\omega_{p}\big)=\frac{1}{FSR}\delta\omega,
\label{eq:S14}
\end{equation}	
where $\delta\omega$ is the detuning $\omega_{0}-\omega_{p}$. The second-order dispersion is given by
\begin{equation}
\delta_{0}=\beta_{1}L\big(\omega_{0}-\omega_{p}\big)=\frac{1}{FSR}\delta\omega,
\label{eq:S15}
\end{equation}	
Eq.~\ref{eq:S12} can be rewritten as  
\begin{equation}
\frac{\partial}{\partial\tau}A=-\left(\frac{\kappa}{2}+i\delta\omega\right)A+i\pi\cdot FSR\cdot D_{2}\frac{\partial^{2}}{\partial T^{2}}A+iL\cdot FSR\cdot\gamma|A|^{2}A+\sqrt{\frac{\kappa\eta P_{in}}{\hbar\omega}}.
\label{eq:S16}
\end{equation}	

The slow time variable $\tau$ in this equation is applied to replace the spatial coordinates $z$. In our topological resonator, the length of the topological resonator is $L=3l$, where $l=180a$ is the side length of the triangular configuration. The FSR and dispersion $D_2$ are extracted from the simulated transmission (Fig. 2(c) in the main text), with calculated values of $FSR=450$ GHz and $D_2=9.53$ GHz. With the Lorentzian fittings of the simulated resonant dips at the pump frequency $\omega_0$, the external loss and the total energy loss rate of the resonator can be calculated as $\kappa_{ex}=1.35\times10^7$ rad/s and $\kappa=2.7\times10^7$ rad/s. The coupling efficiency is given by $\eta=\kappa_{ex}/(\kappa_{ex}+\kappa_{in})=1/2$. The nonlinear coefficient of the topological resonator is given by $\gamma=\omega_{0}n_{2}/cA_{eff}$, where $n_2$ is the nonlinear index of $\rm Si_{3}N_{4}$.

\begin{figure*}[ht]
\centering
\includegraphics[width=1\textwidth]{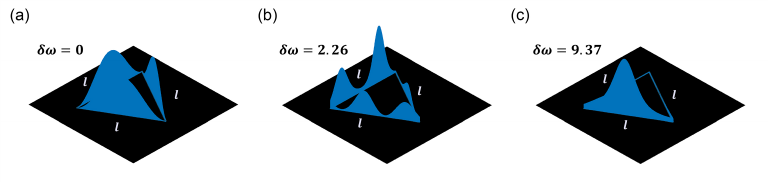}
\caption{Spatial intensity distributions with the detuning of (a) $\delta\omega=0$, (b) $\delta\omega=2.26$, and (c) $\delta\omega=9.37$, respectively.}
    \label{fig:S6}
\end{figure*}

To confirm the spatial distributions of generated solitons, the relative locations of solitons in the topological resonator are related to the propagating time. Since the length of the resonator $L$ corresponds to the roundtrip time T by $L=v_gT$. The corresponding spatiotemporal evolution of the DKS excitation process (Fig. 4(b) in the main text) is a Fourier transform form of frequency-domain combs. In this transform, we set the coupling corner of the triangular resonator as an initiative point, the range distributions of solitons can be calculated from the range of time solutions (0, $L$) directly. Therefore, the relative locations of solitons can be observed in corresponding spatiotemporal evolution. To get deep insight into spatial distributions of the field inside the resonator, we plot the real-time pulse distributions corresponding to four evolutive states inside the topological resonator (Fig. 5 in the main text). And we also show the spatial intensity distributions with the detuning of $\delta\omega=0$, $\delta\omega=2.26$, and $\delta\omega=9.37$, respectively (Fig.~\ref{fig:S6}).

%